%% file: main.tex
\pdfoutput=1
\documentclass[format=acmsmall, review=false, screen=true]{acmart}
\usepackage{multirow}
\usepackage{amsthm,amsmath,nccmath}
\usepackage{caption}
\usepackage{xcolor}
\usepackage{subfigure}
\usepackage{wrapfig}
\usepackage{listings}
\usepackage[ruled, vlined, linesnumbered]{algorithm2e}
\usepackage{algpseudocode}
\usepackage{enumitem}
\usepackage{csquotes}
\usepackage{balance}
\usepackage{multicol}
\usepackage{makecell}
\usepackage{marvosym}
\usepackage{tcolorbox}
\usepackage{soul}

\definecolor{purple}{RGB}{236,236,252}
\definecolor{dkblue}{RGB}{0,0,123}
\newcommand\ans[1]{
\begin{tcolorbox}[boxrule=0pt, colback=purple, colframe=white, size=title]
\noindent #1
\end{tcolorbox}
}

\def \tool {{\sc Synthify}\xspace}
\def \toolbf {{\sc \textbf{Synthify}}\xspace}
\def \baseline {{\sc PSY-TaLiRo}\xspace}

\SetKwInput{KwInput}{Require}
\SetEndCharOfAlgoLine{}

\definecolor{codegreen}{rgb}{0,0.6,0}
\definecolor{codegray}{rgb}{0.5,0.5,0.5}
\definecolor{codepurple}{RGB}{117, 20, 124}
\definecolor{backcolour}{rgb}{0.95,0.95,0.92}
\lstdefinestyle{mystyle}{
  language=Python,
  backgroundcolor=\color{backcolour}, commentstyle=\color{codegreen},
  keywordstyle=\color{codepurple},
  numberstyle=\color{codegray},
  stringstyle=\color{codepurple},
  basicstyle=\ttfamily\small,
  breakatwhitespace=false,
  breaklines=true,
  captionpos=t,
  keepspaces=true,
  numbers=left,
  numbersep=2pt,
  showspaces=false,
  showstringspaces=false,
  showtabs=false,
  tabsize=1
}

\AtBeginDocument{%
  \providecommand\BibTeX{{%
    \normalfont B\kern-0.5em{\scshape i\kern-0.25em b}\kern-0.8em\TeX}}}

\setcopyright{acmlicensed}
\acmJournal{TOSEM}

\begin{document}

\title{Finding Safety Violations of AI-Enabled Control Systems through the Lens of Synthesized Proxy Programs}

\author{Jieke Shi}
\email{jiekeshi@smu.edu.sg}
\orcid{0000-0002-0799-5018}
\author{Zhou Yang}
\authornote{Corresponding author.}
\orcid{0000-0001-5938-1918}
\email{zyang@smu.edu.sg}
\author{Junda He}
\email{jundahe@smu.edu.sg}
\orcid{0000-0003-3370-8585}
\affiliation{%
  \institution{Singapore Management University}
  \country{Singapore}
}

\author{Bowen Xu}
\email{bxu22@ncsu.edu}
\orcid{0000-0002-1006-8493}
\affiliation{%
  \institution{North Carolina State University}
  \country{United States}
}

\author{Dongsun Kim}
\email{darkrsw@korea.ac.kr}
\orcid{0000-0003-0272-6860}
\affiliation{%
  \institution{Korea University}
  \country{South Korea}
}

\author{DongGyun Han}
\email{DongGyun.Han@rhul.ac.uk}
\orcid{0000-0002-8599-2197}
\affiliation{%
  \institution{Royal Holloway, University of London}
  \country{United Kingdom}
}

\author{David Lo}
\email{davidlo@smu.edu.sg}
\orcid{0000-0002-4367-7201}
\affiliation{%
  \institution{Singapore Management University}
  \country{Singapore}
}

\ccsdesc[500]{Software and its engineering~Software testing and debugging}
\ccsdesc[500]{Software and its engineering~Search-based software engineering}
\ccsdesc[500]{Computing methodologies~Control methods}

\begin{abstract}
  Given the increasing adoption of modern AI-enabled control systems, ensuring their safety and reliability has become a critical task in software testing. One prevalent approach to testing control systems is falsification, which aims to find an input signal that causes the control system to violate a formal safety specification using optimization algorithms. However, applying falsification to AI-enabled control systems poses two significant challenges: (1)~it requires the system to execute numerous candidate test inputs, which can be time-consuming, particularly for systems with AI models that have many parameters, and (2)~multiple safety requirements are typically defined as a conjunctive specification, which is difficult for existing falsification approaches to comprehensively cover.

  This paper introduces \tool, a falsification framework tailored for AI-enabled control systems, i.e., control systems equipped with AI controllers. Our approach performs falsification in a two-phase process. At the start, \tool synthesizes a program that implements one or a few linear controllers to serve as a proxy for the AI controller. This proxy program mimics the AI controller's functionality but is computationally more efficient. Then, \tool employs the $\epsilon$-greedy strategy to sample a promising sub-specification from the conjunctive safety specification. It then uses a Simulated Annealing-based falsification algorithm to find violations of the sampled sub-specification for the control system. To evaluate \tool, we compare it to \baseline, a state-of-the-art and industrial-strength falsification tool, on 8 publicly available control systems. On average, \tool achieves a 83.5\% higher success rate in falsification compared to \baseline with the same budget of falsification trials. Additionally, our method is 12.8$\times$ faster in finding a single safety violation than the baseline. The safety violations found by \tool are also more diverse than those found by \baseline, covering 137.7\% more sub-specifications.
\end{abstract}

\keywords{Falsification, Search-based Testing, AI-enabled Control Systems, Program Synthesis}

\maketitle

\input{sections/intro.tex}

\input{sections/preliminary.tex}
\input{sections/motivation.tex}
\input{sections/methodology.tex}
\input{sections/experiments.tex}
\input{sections/results.tex}
\input{sections/discussion.tex}
\input{sections/rel_work.tex}
\input{sections/conclusion.tex}

\begin{acks}
  This research/project is supported by the National Research Foundation Singapore and DSO National Laboratories under the AI Singapore Programme (AISG Award No: AISG2-RP-2020-017). This work was also supported by the Institute of Information \& Communications Technology Planning \& Evaluation (IITP)--ICT Creative Consilience Program grant funded by the Korea government (MSIT) (IITP-2025-RS-2020-II201819). This work was also supported by the National Research Foundation of Korea (NRF) grant funded by the Korea government (MSIT) (No. 2021R1I1A3048013).
\end{acks}

\balance
\bibliographystyle{ACM-Reference-Format}
\bibliography{ref}

\end{document}

%% file: sections/intro.tex
\section{Introduction}
\label{sec:intro}

Control systems, which utilize software controllers to maintain desired behaviors of particular mechanical systems, have been broadly deployed into diverse real-world applications. These systems have become an essential part of everyday life and industrial production~\cite{XiaGYF15}. With the rapid advances of Artificial Intelligence (AI) in recent decades, there has been a huge demand for enhancing or even replacing traditional mathematical controllers (e.g., Proportional-Integral-Derivative~\cite{johnson2005pid}) with AI controllers (e.g., Deep Neural Networks) to achieve more optimized performance. We refer to control systems employing AI controllers as AI-enabled control systems. With their immense potential, AI-enabled control systems have inevitably been or will soon be adopted in various safety-critical domains, such as aerospace~\cite{izzo2019survey,chai2021review}, autonomous driving~\cite{kiran2021deep, muhammad2020deep}, healthcare monitoring~\cite{kishor2022artificial,kamruzzaman2021new}, and even nuclear fusion~\cite{degrave2022magnetic}. Therefore, ensuring the reliability and safety of AI-enabled control systems in such contexts is of paramount importance.

To date, falsification has been shown to be an effective approach for testing complex control systems~\cite{song2022cyber,corso2021survey,menghi2020approximation,yaghoubi2017hybrid,zhang2022falsifai,zhang2021figcps}. Concretely, given a control system with a safety specification, usually written in Signal Temporal Logic (STL)~\cite{maler2004monitoring}, falsification aims to find an input signal that causes the control system to violate the specification. Falsification often uses an optimization algorithm to search for input signals with the goal of minimizing the quantitative robustness semantics of temporal logics~\cite{donze2010robust}. These semantics quantify the degree to which an input signal deviates from violating the safety specification as a real number. A smaller number means that the corresponding input signal is more likely to violate a given safety specification, and a negative number suggests that the safety specification is violated. Falsification iteratively generates candidate input signals and executes them on the control system until an input with negative robustness semantics is found, revealing a violation of the safety specification.
Various falsification methods have been proposed and adopted in practice~\cite{eddeland2020industrial,tuncali2018experience,kapinski2016simulation}, showing that falsification is effective in revealing specification violations in complex control systems.

However, applying falsification in AI-enabled control systems poses two challenges. First, falsification of AI-enabled control systems suffers from the scalability issue. Falsification requires the system under test to execute numerous candidate inputs until a safety violation is found. However, AI controllers are typically implemented as deep neural networks (DNNs) with numerous parameters, which are computationally expensive to execute. For example, DNN-based controllers in~\cite{zhu2019inductive} have 4--50$\times$ longer execution time than classic linear controllers. Consequently, falsification can take a large amount of time to the extent that it becomes impractical.

Second, falsification of AI-enabled control systems fails to comprehensively cover conjunctive specifications. AI-enabled control systems emerge to handle highly complex control tasks, often with multiple safety specifications. For instance, a self-driving control system in~\cite{zhu2019inductive} is required to maintain a car's heading angle within $90^{\circ}$ ($\varphi_1$) while keeping the distance between the car and the centerline of the road below $2.0$ meters ($\varphi_2$). A common practice for falsifying such multiple specifications simultaneously is to combine them as a conjunctive specification ($\varphi_1 \wedge \varphi_2$)~\cite{liden2022multi}. However, directly falsifying such a conjunctive specification often produces less diverse safety violations. If one sub-specification is easily violated, the falsification process tends to repeatedly exploit that sub-specification to report violations and overlook others. As Sun et al.~\cite{sun2022lawbreaker} discussed, it is desirable to find diverse violations that can cover as many sub-specifications as possible. Nevertheless, the existing state-of-the-art falsification tools like \baseline~\cite{thibeault2021psy} are not inherently designed to meet this challenge.

Inspired by recent advances in controller synthesis~\cite{mania2018simple,zhu2019inductive}, our key insight is that a complex AI controller can be approximated into a program that implements one or a few linear controllers, which is less computationally expensive to run. Of course, such a program with linear controllers may not precisely mirror the AI controller's performance, but it does achieve similar functionality to some extent~\cite{mania2018simple}. For example, experiments by Rajeswaran et al.~\cite{NIPS2017_9ddb9dd5} show that robots controlled by linear controllers may not walk like AI-enabled ones, but they can still stand up effectively. As analyzed by Zhu et al.~\cite{zhu2019inductive}, a program having similar functionality to an AI controller can also express similar safety properties. Thus, we propose to mitigate the scalability issue in the falsification of AI-enabled control systems by synthesizing computationally cheap programs as proxies for AI controllers.

Additionally, we consider that recent studies have highlighted the importance of the exploration-exploitation trade-off in falsifying conjunctive specifications~\cite{mathesen2021efficient,ramezani2022falsification}. Specifically, falsification can either exploit a sub-specification that is easily violated to ensure successful falsification, or explore a sub-specification that has not yet been falsified to boost the coverage of all sub-specifications within the conjunctive specification. The low coverage of sub-specifications is often a result of inadequate exploration and over-exploitation. Thus, we propose a strategy that balances the exploration and exploitation of sub-specifications, allowing falsification to more comprehensively cover the conjunctive specification.

Pursuant to the above insights, we propose a novel falsification framework named \tool ({\sc Syn}thesize-{\sc th}en-fals{\sc ify}). \tool is tailored for AI-enabled control systems using Deep Reinforcement Learning (DRL) algorithms, which are considered as the state-of-the-art in various control tasks. To facilitate efficient and effective falsification of AI-enabled control systems,
\tool combines two distinct techniques: (1)~sketch-based program synthesis~\cite{solar2009sketching} to effectively synthesize a proxy program from the DRL-based AI controller. The proxy program implements one or a few linear controllers to mimic the functionality of the AI controller but is much cheaper to run; (2)~$\epsilon$-greedy strategy~\cite{kuleshov2014algorithms}, a classic strategy for balancing exploration and exploitation in the decision-making process. This strategy helps \tool sample a promising sub-specification from the given conjunctive safety specification. \tool then utilizes the synthesized program as a proxy for the AI controller in the control system and employs a Simulated Annealing-based algorithm to guide the search for violations of the sampled sub-specification.

We evaluate \tool on 8 publicly available control systems from the literature~\cite{zhu2019inductive} and compare it to a state-of-the-art falsification tool \baseline~\cite{thibeault2021psy}, which recently won the 2022 ARCH-COMP prize for its technical achievements and excellent performance in falsification~\cite{ernst2022arch}. Our results show that, given the same budget of falsification trials, \tool achieves a 83.5\% higher falsification success rate than \baseline, while also being 5.6$\times$ faster than the baseline at completing the same number of falsification trials. Given the same time budget, \tool reveals 7.8$\times$ more violations of safety specifications than \baseline. Moreover, our method is 12.8$\times$ faster than the baseline at finding one single violation. The safety violations found by \tool are also more diverse than those found by \baseline, improving 137.7\% the coverage of the conjunctive specification on average. These results demonstrate the effectiveness and efficiency of the proposed method in testing the safety of AI-enabled control systems.

The contributions of this paper are summarized as follows:
\begin{itemize}
    \item {\bf Insight:} We are the first to combine sketch-based program synthesis and $\epsilon$-greedy strategy to simultaneously address the scalability and sub-specification coverage issues in the falsification of AI-enabled control systems.
    \item {\bf Technique:} We propose and implement \tool, an innovative falsification framework that can both efficiently and effectively uncover safety violations of AI-enabled control systems.
    \item {\bf Evaluation:} We evaluate \tool on 8 control systems and compare it to \baseline, a state-of-the-art falsification tool. The results show that our method outperforms the baseline by a large margin in terms of both effectiveness and efficiency.
\end{itemize}

The remainder of the paper is organized as follows. Section~\ref{sec:backgrounds} covers the preliminary information of our work. Section~\ref{sec:motivation} shows an example to motivate our work, while Section~\ref{sec:approach} describes in detail how \tool works. Section~\ref{sec:exp_setup} presents the experimental setup and research questions. Section~\ref{sec:results} analyzes the experimental results. Section~\ref{sec:discussion} presents a number of additional analyses and discussions. Section~\ref{sec:rel_work} presents related work, and Section~\ref{sec:conclusion} concludes our paper and provides information about our replication package.

%% file: sections/preliminary.tex
\section{Preliminaries}
\label{sec:backgrounds}

This section provides an introduction to AI-enabled control systems. It also gives a brief description of Signal Temporal Logic (STL) as a formal specification language for control systems, and falsification, a common and important testing method used in the safety analysis of control systems.

\begin{figure}
    \centering
    \includegraphics[width=0.8\linewidth]{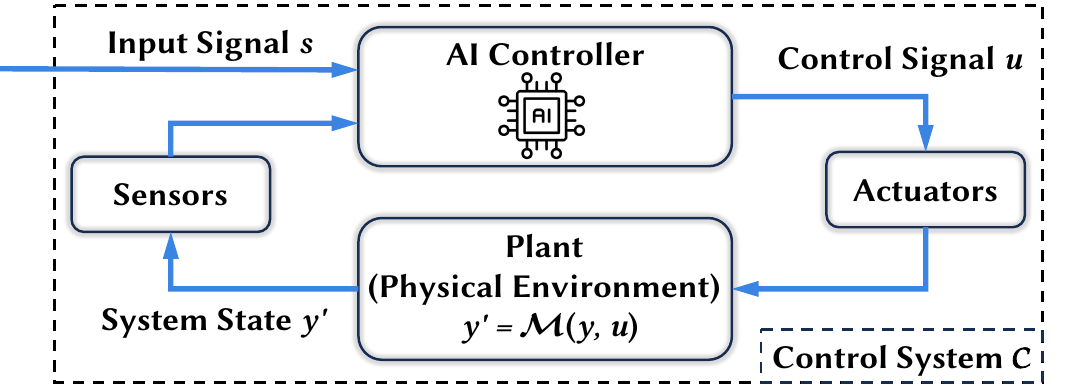}
    \caption{Architecture of an AI-enabled control system.}
    \label{fig:control-systems}
\end{figure}

\subsection{AI-enabled Control System}
Control systems utilize software controllers to regulate the behavior of mechanical systems based on real-time feedback from their physical environments~\cite{zhang2017analysis}. Traditionally, controllers were designed with classic algorithms like Proportional-Integral-Derivative (PID)~\cite{johnson2005pid}. However, with the growing interest in AI, there has been an exploration of AI controllers powered by deep neural networks. We refer to control systems equipped with AI controllers as AI-enabled control systems.

Figure~\ref{fig:control-systems} shows a typical architecture of AI-enabled control systems consisting of a plant, a controller, sensors, and actuators. The plant $\mathcal{M}$ is a physical environment whose next state $y^\prime$ is determined by the current state $y$ and the control signal $u$. The control signal $u$ is issued by an AI controller and applied to the plant by actuators. Integrating AI controllers allows control systems to adapt to more complex situations and often achieve superior results compared to traditional methods~\cite{hunt1992neural, LillicrapHPHETS15}. However, this integration also introduces new risks, such as the potential for system failures that could violate safety requirements. These risks underscore the need for our work in analyzing the safety and reliability of AI-enabled control systems.

The AI controller is typically implemented as a deep neural network, and Deep Reinforcement Learning (DRL) is currently considered the state-of-the-art method for training it~\cite{song2022cyber,duan2016benchmarking,LillicrapHPHETS15}. Our study focuses on control systems with DRL-based AI controllers. DRL is a subfield of machine learning that focuses on learning optimal decision strategies through interactions with the environment~\cite{wang2022deep}. In DRL, an agent learns the best policy by observing states and selecting the best output to maximize the cumulative reward. Concretely, considering a DRL-based AI controller as an agent, it receives the current state of the plant and outputs a control signal to transition the system to a new state. A policy, a deep neural network taking the system's state as input, determines the control signal that the agent outputs. The reward for each output under a state depends on the feedback from the plant. Through repeated interactions with the plant, the agent is trained to maximize the reward it receives, learning and adapting its policy to make better decisions over time. After training, the agent with its best policy is deployed as an AI controller in the control system. DRL-based AI controllers have found applications in many critical areas such as autonomous driving~\cite{kiran2021deep,wu2021deep}, robotics control~\cite{ju2022transferring, hester2012rtmba}, etc.

\subsection{Signal Temporal Logic}
The safety requirements of a control system are usually specified using Signal Temporal Logic (STL)~\cite{maler2004monitoring}, a widely-adopted formal specification language. STL describes how a control system should behave by combining atomic propositions and formulas using logical and temporal operators. Formally, the basic syntax of an STL formula $\varphi$ is defined as:
\begin{equation}
    \varphi ::= \mu \ \vert \ \bot \ \vert \ \neg \varphi \ \vert \ \varphi_1 \wedge \varphi_2 \ \vert \ \varphi_1 \ \mathcal{U}_{\mathcal{I}} \ \varphi_2
\end{equation}
where $\mu ::= f(s) \sim c$ is an atomic proposition, $f(s)$ is a real-valued function over the variables $s$, $\sim \in \{<, \le, =, \ge, >\}$ is a comparison operator, and $c$ is a constant. Additionally, $\bot$ represents ``false'', while $\neg$ and $\wedge$ are the negation and conjunction operators, respectively. $\mathcal{U}_{\mathcal{I}}$ is the ``until'' operator with a time interval $\mathcal{I}$, which is a closed non-singular interval in $\mathbb{R}_{\ge 0}$, i.e., $\mathcal{I} = [a, b]$ or $[a, \infty]$ where $a, b \in \mathbb{R}$ and $a < b$. From the basic definition, we can derive other logical and temporal operators, including $\top \equiv \neg \bot$ (true), $\vee \equiv \neg(\neg \varphi_1 \wedge \neg \varphi_2)$ (disjunction), $\rightarrow \equiv \neg \varphi_1 \vee \varphi_2$ (implication), $\Diamond_{\mathcal{I}} \varphi \equiv \top \ \mathcal{U}_{\mathcal{I}} \ \varphi$ (eventually), and $\square_{\mathcal{I}} \varphi \equiv \neg \Diamond_{\mathcal{I}} \neg \varphi$ (always).

STL has quantitative robustness semantics~\cite{donze2010robust} that map an input signal $s$ and an STL specification $\varphi$ to a real number, indicating the extent to which $s$ satisfies $\varphi$. Formally, with respect to a signal $s = (s_1, s_2, ..., s_n)$, the satisfaction of an atomic proposition $\mu = f(s) \sim c$ at time $t$ is defined as:
\begin{equation}
    \small
    \begin{array}{lll}
    (s, t) \models \mu & \Leftrightarrow & f\left(s_1[t], \ldots, s_n[t]\right) \sim c \\
    (s, t) \models \varphi_1 \wedge \varphi_2 & \Leftrightarrow & (s, t) \models \varphi_1 \wedge(s, t) \models \varphi_2 \\
    (s, t) \models \neg \varphi & \Leftrightarrow & \neg((s, t) \models \varphi) \\
    (s, t) \models \varphi \mathcal{U}_{\mathcal{I} = [a, b]} \varphi & \Leftrightarrow & \exists t^{\prime} \in[t+a, t+b] \text {\it \ such that }\left(s, t^{\prime}\right) \models \varphi \wedge \forall t^{\prime \prime} \in\left[t, t^{\prime}\right],\left(s, t^{\prime \prime}\right) \models \varphi \\
    (s, t) \models \Diamond_{\mathcal{I} = [a, b]} \varphi & \Leftrightarrow & \exists t^{\prime} \in[t+a, t+b] \text {\it \ such that }\left(s, t^{\prime}\right) \models \varphi \\
    (s, t) \models \square_{\mathcal{I} = [a, b]} \varphi & \Leftrightarrow & \forall t^{\prime} \in[t, t+b] \text {\it \ such that }\left(s, t^{\prime}\right) \models \varphi
    \end{array}
\end{equation}
The larger the semantic value, the more robustly the input signal $s$ satisfies the STL formula $\varphi$. A negative value indicates that $s$ violates $\varphi$, with the degree of violation equal to the absolute value of the number. The intuition behind the robustness semantics is to measure the distances of the points on the system trace from the boundaries of the sets defined by the formula predicates using maximum or minimum functions. For example, let us consider the STL formula $\square_{[0, 1]}(a < 1.0)$, specifying that a variable $a$ must remain below $1.0$ for the duration of the time interval $[0, 1]$. Suppose we have a system trace $[(0.5, 0), (1.2, 1)]$, where each element $(a, time)$ corresponds to $a$'s value and its associated time. For the first point, the value of $a$ is below the threshold $1.0$, so the distance to the boundary is simply the difference between the threshold and the value: $1.0 - 0.5 = 0.5$. For the second point, $a$ goes up to the threshold, so the distance to the boundary is $1.0 - 1.2 = -0.2$. To compute the robustness semantics of the trace, we take the minimum distance across all points within the time interval, which in this case is a negative value of $-0.2$, indicating that the signal $s$ violates the STL formula by $0.2$. We refer interested readers to~\cite{donze2010robust} for the complete definition of STL robustness semantics.

\subsection{Falsification}
Falsification is a well-established testing method used in the safety analysis of control systems. It aims to find inputs to a control system that cause the system to violate safety specifications expressed in STL. Concretely, given a control system $\mathcal{C}$ and an STL specification $\varphi$, falsification first searches for an input signal $s$ and executes the system to obtain a system state trace. The trace is then used to compute the quantitative robustness semantics with respect to $\varphi$. If negative robustness is obtained, indicating a violation of $\varphi$, the falsification is successful and the input signal $s$ is considered a falsifying input to the control system $\mathcal{C}$. The search continues until a violation is found or a given budget (time or number of executions) is exhausted. Notably, falsification is typically performed in a black-box manner, i.e., the internal structures of the plant and the AI controller are unknown throughout the falsification process; only the input and the corresponding output signals can be observed.

\begin{algorithm}[!t]
    \small
    \SetKwInput{Input}{Require}
    \SetKwInput{Para}{Parameters}
    \SetKwComment{Comment}{/* }{ \texttt{*/}}
    \SetKwFunction{Select}{{\sc RandomSelect}}
    \SetKwFunction{FRobust}{{\sc RobustSemantics}}
    \SetKwFunction{FPerturb}{{\sc Perturb}}
    \SetKwFunction{FRandom}{\sc Random}
    \Input{Control system $\mathcal{C}$, STL specification $\varphi$, Input ranges $\mathcal{S}$}
    \Para{Budget $\mathcal{B}$, Maximum number of iterations $N$, Cooling rate $\alpha$, and Initial temperature $T$}

    \Repeat{\textup{\sc Budget $\mathcal{B}$ (time or number of trials) is exhausted}}
    {
        $s \gets \Select(\mathcal{S})$ \label{algo:SA:select}  \Comment*[r]{\it Randomly initialize $s$ from the given input ranges $\mathcal{S}$}
        $rb \gets \FRobust(\mathcal{C}(s), \varphi)$ \\
        \lIf{$rb < 0$}{ \Return $s$ \Comment*[f]{\it Return $s$ if getting a negative robustness}}
        \For(\Comment*[f]{\it Simulated Annealing with $N$ iterations}){$i \gets 1$ \KwTo $N$}{
            $s^\prime \gets$ \FPerturb($s$) \label{algo:SA:perturb} \\
            $rb_{s^\prime} \gets \FRobust(\mathcal{C}(s^\prime), \varphi)$ \label{algo:SA:robust} \\
            \lIf{$rb_{s^\prime} < 0$}{ \Return $s^\prime$ \Comment*[f]{\it Return $s^\prime$ if getting a negative robustness}} \label{algo:SA:return}
            $\Delta \gets rb_{s^\prime} - rb$, $P \gets e^{-\Delta/T} $ \label{algo:SA:delta} \Comment*[r]{\it Accept $s^\prime$ if it is better or by probability }
            \If{$\Delta < 0$ or \  $\FRandom(0, 1) < P$}{
                $s \gets s^\prime$, $rb \gets  rb_{s^\prime}$ \label{algo:SA:accept}
            }
            $T \gets \alpha*T$\Comment*[r]{\it Update temperature}
        }
    }
    \Return{\textup{\sc None}}\Comment*[r]{\it No input that causes violations found}
    \caption{Simulated Annealing-based Falsification}
    \label{algo:SA}
\end{algorithm}

Falsification is typically supported by optimization algorithms such as Simulated Annealing (SA)~\cite{abbas2012convergence} shown in Algorithm~\ref{algo:SA}. The algorithm starts with a randomly-initialized input signal $s$ (line~\ref{algo:SA:select}). It then iteratively perturbs it to generate a new input $s^\prime$ (line~\ref{algo:SA:perturb}), which is then used to execute the system $\mathcal{C}$ to check if it is a falsifying input (lines~\ref{algo:SA:robust}-\ref{algo:SA:return}). If $s^\prime$ does not cause the system to violate $\varphi$, i.e., obtaining non-negative robust semantics, the algorithm accepts $s^\prime$ into the next iteration with a probability that is determined by the temperature $T$ (lines~\ref{algo:SA:delta}-\ref{algo:SA:accept}). The algorithm terminates when the budget $\mathcal{B}$ is exhausted or when a falsifying input signal $s$ is found, witnessing the violation of $\varphi$.


\subsection{$\epsilon$-greedy Strategy}
\label{subsec:epsilon-greedy}

When iteratively selecting among multiple options, a key challenge is to balance exploration (i.e., trying new options that might yield better results) and exploitation (i.e., choosing the best option based on current knowledge). In this context, the $\epsilon$-greedy strategy~\cite{sutton2018reinforcement} offers a straightforward yet effective solution. The $\epsilon$ in its name represents a probability that governs the trade-off between exploration and exploitation. Formally, consider a set of $n$ options $O = \{o_1, o_2, \dots, o_n\}$, where the performance of each option is evaluated through reward feedback, we can obtain the knowledge of the expected reward $\mathcal{R} = \{r_1, r_2, \dots, r_n\}$ of each option by averaging the rewards received from multiple attempts so far. Then, as shown in Algorithm~\ref{algo:epsilon-greedy1}, with a probability of $1 - \epsilon$, the strategy exploits the current knowledge by selecting the option with the highest expected reward (line~\ref{algo:epsilon-greedy1:exploit}). Alternatively, with probability $\epsilon$, it explores by selecting an option at random (line~\ref{algo:epsilon-greedy1:random}), enabling the discovery of potentially better alternatives.

\begin{algorithm}[!t]
    \small
    \caption{$\epsilon$-greedy Strategy}
    \label{algo:epsilon-greedy1}
    \SetKwComment{Comment}{/* }{ */}
    \SetKwInput{Input}{Require}
    \SetKwInput{Para}{Parameters}
    \Input{Options $O = \{o_1, o_2, ..., o_n\}$, Expected rewards $\mathcal{R} = \{r_1, r_2, ..., r_n\}$}
    \Para{Probability threshold $\epsilon$}
    \SetKwFunction{FRandom}{\sc Random}
    \SetKwFunction{FRandomS}{\sc RandomSample}
    \eIf{$\FRandom(0, 1) < \epsilon$\label{algo:epsilon-greedy1:random}}{
        $o \gets \FRandomS(O)$ \Comment*[r]{\it Randomly select an option}
    }
    {
        $o \gets \arg\max_{o_i \in O} \mathcal{R}_i, i \in [1, n]$ \label{algo:epsilon-greedy1:exploit} \Comment*[r]{\it Select the option with the highest expected reward}
    }
    \Return $o$ \Comment*[r]{\it Return the selected option}
\end{algorithm}

As shown by Zhang et al.~\cite{zhang2021effectiveness}, certain sub-specifications within a conjunctive specification are significantly easier to falsify than others as they are more sensitive to input signals. This sensitivity causes falsification algorithms to disproportionately focus on these easier sub-specifications, reducing the likelihood of falsifying more difficult ones and potentially resulting in an incomplete safety analysis. Our approach treats each sub-specification as an independent option and incorporates the $\epsilon$-greedy strategy to encourage the falsification of sub-specifications that have not yet been falsified, thereby ensuring a more comprehensive safety analysis. A detailed explanation of how the $\epsilon$-greedy strategy works in our approach is provided in Section~\ref{subsec:sampling}.

%% file: sections/motivation.tex
\section{Motivating Example}
\label{sec:motivation}

This section uses the example of a self-driving system provided by Zhu et al.~\cite{zhu2019inductive} to better illustrate the two main challenges associated with the falsification of AI-enabled control systems. We then show how \tool effectively addresses these challenges.

\begin{figure}[t!]
    \centering
    \includegraphics[width=0.25\linewidth]{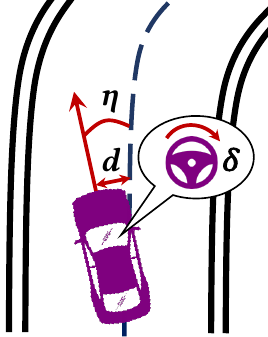}
    \caption{An example of self-driving control system.}
    \label{fig:self-driving}
\end{figure}

\vspace{0.3em}
\noindent\textbf{Self-Driving System.}\hspace{4pt} Figure~\ref{fig:self-driving} presents a self-driving control system that uses a DRL-based AI controller to control a car to follow a given route. The system has a 2-dimensional state $[\eta, d]$, where $\eta$ is the angle between the car's heading and the road's centerline, and $d$ is the distance between the car's front and centerline. The AI controller takes the state as input and outputs a 1-dimensional control signal $\delta$, which is used to control the steering angle. A regulator ensures the signal $\delta$ remains within the range of $[-10, 10]$, preventing rapid or extreme steering adjustments.
As described by Zhu et al.~\cite{zhu2019inductive}, the car keeps moving at a constant speed, and the safety property of this system is to ensure that the car never veers into canals found on both sides of the road. Formally, the self-driving system is safe if the car's heading angle $\eta$ remains within a threshold of $90^{\circ}$ and the distance $d$ between the car's front and the road's centerline stays below $2.0$ meters. Such safety requirements are defined as the following conjunctive specification written in STL:
\begin{equation}
    \label{eqn:spec}
    \varphi_{safety} \equiv  \square_{[0, 200]} \left( \vert\eta\vert <90^{\circ} \wedge\vert d \vert < 2.0 \right)
\end{equation}
where the temporal operator $\square_{[0, 200]}$ indicates that the safety property must hold continuously for 200 time steps throughout the system's execution. This control system, though simple, serves as a representative example of many practical control systems that utilize AI models to govern the system's behavior, such as the Udacity self-driving car~\cite{Udacity}. We employ this example to clearly illustrate specific challenges that motivate the design of \tool, while more complex control systems are provided in Section~\ref{sec:benchmarks}.

\begin{table}[t!]
    \centering
    \caption{Running time and sub-specification violation coverage of the SOTA falsification tool \baseline on the self-driving system over the 50 trials.}
    \label{tab:motivation}
    \renewcommand{\familydefault}{\sfdefault}\normalfont
    \small
    \begin{tabular}{c|c|c||c}
        \hline\hline
        \multicolumn{3}{c||}{Running time} & \multirow{2}{*}{\begin{tabular}[c]{@{}c@{}}Sub-specification\\ Coverage\end{tabular}} \\
        \cline{1-3}
        AI Controller & Plant Execution & Algorithm & \\
        \hline
        \textbf{408.7s / 78.8\%} & 73.8s / 14.2\% & 36.1s / 7.0\% & \textbf{50.0\%}\\
        \hline\hline
    \end{tabular}
\end{table}

\vspace{0.3em}
\noindent\textbf{Key Challenges.}\hspace{4pt}
We use the state-of-the-art falsification tool, \baseline~\cite{thibeault2021psy}, which employs the Simulated Annealing-based falsification algorithm outlined in Algorithm~\ref{algo:SA}, to find safety violations of the self-driving system. We set the maximum number of falsification trials to 50 and feed the above-mentioned safety specification into \baseline.
The results of running time and sub-specification violation coverage are presented in Table~\ref{tab:motivation}. Following Sun et al.~\cite{sun2022lawbreaker}, the sub-specification violation coverage is defined as follows:

\vspace{0.3em}
\noindent\textit{\textbf{Definition 1:} Sub-specification Violation Coverage.}  Given a set of violations $\mathcal{V}$ against a safety specification $\varphi$ having $n$ sub-specifications, the sub-specification violation coverage is the ratio of violated sub-specifications to the total number of sub-specifications, i.e.,
\begin{equation}
    \text{Sub-specification Coverage} = \frac{\left| \{ \varphi_i \mid \exists v \in \mathcal{V}, \neg(v \models \varphi_i), i \in [1, n] \} \right|}{n} \times 100\%
\end{equation}

From Table~\ref{tab:motivation}, we can see that in the 50 falsification trials, the execution of the AI controller is the most time-consuming part, taking 408.7 seconds, which accounts for 78.8\% of the total running time. This significant duration is attributed to the AI controller's numerous parameters, resulting in high execution costs. This in turn makes the overall falsification time too long to be practical. Thus, the first challenge in falsification of AI-enabled control systems concerns the scalability issue, which can be addressed by reducing the cost of executing the AI controllers.

Additionally, in the example of the self-driving system, the safety requirements in Equation~\ref{eqn:spec} is a conjunctive specification, which can be divided into four sub-specifications as follows:
\begin{equation}
    \small
    \label{eqn:subspec}
    \begin{gathered}
    \varphi_{safety} \equiv\left\{\varphi_1 \equiv \square_{[0,200]}\left(\eta>-90^{\circ}\right), \varphi_2 \equiv \square_{[0,200]}\left(\eta<90^{\circ}\right),\right. \\
    \left.\varphi_3 \equiv \square_{[0,200]}(d>-2.0), \varphi_4 \equiv \square_{[0,200]}(d<2.0)\right\}
    \end{gathered}
\end{equation}
We then calculated the sub-specification violation coverage of \baseline, as shown in Table~\ref{tab:motivation}. Within 50 trials, the safety violations found by \baseline could only falsify 50\% of the sub-specifications within the conjunctive specification, specifically the first 2 of the 4 sub-specifications in Equation~\ref{eqn:subspec}. The remaining 2 sub-specifications were missed by \baseline. This result suggests that another challenge in falsification of AI-enabled control systems is the lack of comprehensiveness in covering all the sub-specifications required for ensuring safety.

\vspace{0.3em}
\noindent\textbf{How \& Why \toolbf Works.}\hspace{4pt}
Our \tool is purpose-built to address the above two challenges in the falsification of AI-enabled control systems. To improve scalability, we propose an Evolution Strategy-based synthesis algorithm (Section~\ref{subsec:synthesis}), which allows \tool to synthesize a program to act as a proxy for the AI controller in the control system. To offer a clearer insight into the synthesized proxy program, we present an example in the self-driving scenario:
\begin{lstlisting}[mathescape]
  def self_driving($\eta$, $d$):
      $\delta$ = $0.20706786 * \eta$ - $0.31286586 * d$ - $0.27174068$
      return $\delta$
\end{lstlisting}
The proxy program includes a linear controller of the system states $\eta$ and $d$ and return a control signal $\delta$ to control the car's steering angle. It closely emulates the control behavior of the original AI controller, while being much more efficient to execute. During the falsification process, the proxy program replaces the original AI controller in the control system, and the AI controller is only used to verify if a real safety violation has been found. As a result, the execution cost of the AI controller is considerably reduced, effectively mitigating the scalability issue.

Furthermore, to enhance the sub-specification violation coverage, \tool proposes an $\epsilon$-greedy strategy-based method, which involves sampling a promising sub-specification from the conjunctive specification for the subsequent falsification process. The $\epsilon$-greedy strategy incorporates a probability of randomly selecting a sub-specification, which gives a chance for the falsification process to explore sub-specifications that have not been falsified yet, significantly improving the sub-specification coverage. More information on the $\epsilon$-greedy strategy can be found in Section~\ref{subsec:sampling}.

%% file: sections/methodology.tex
\section{Detailed Approach}
\label{sec:approach}

\begin{figure}[!t]
    \centering
    \includegraphics[width=0.8\linewidth]{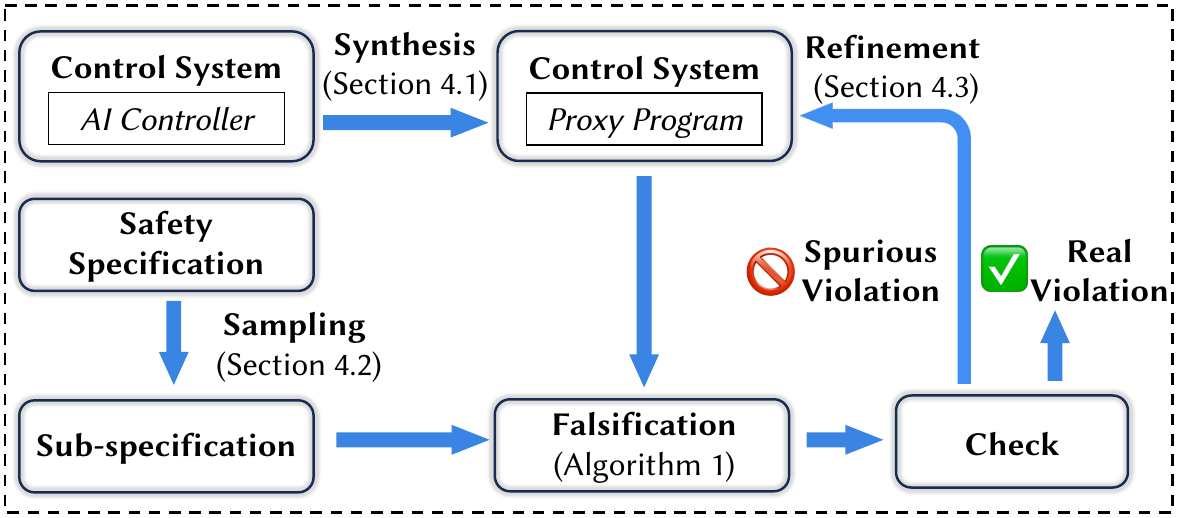}
    \caption{Overview of {\sf \textsc{Synthify}}.}
    \label{fig:overview}
\end{figure}

This section provides a detailed introduction to \tool. Figure~\ref{fig:overview} shows an overview of our approach. Given a control system with an AI controller, the approach first approximates a proxy program from the controller to replace the controller (Section~\ref{subsec:synthesis}). Next, we separate the conjunctive safety specifications and sample one from them as the input to the subsequent falsification algorithm (Section~\ref{subsec:sampling}). Finally, we employ the Simulated Annealing-based algorithm to search for violations of the sampled specification (Algorithm~\ref{algo:SA}). If a violation is detected in the control system with the proxy program, we check whether it is spurious, i.e., it does not reveal any violations in the original control system. If spurious, we use it to refine the proxy program (Section~\ref{subsec:refinement}); otherwise, we report it as a violation of safety specifications.

\subsection{Synthesis}
\label{subsec:synthesis}
To facilitate the efficient falsification of AI-enabled control systems,
\tool first synthesizes a proxy program for the AI controller that is computationally more efficient.
The proxy program replaces the AI controller in the control system during falsification.
Our synthesis procedure consists of two steps: (1) constructing a sketch that can express the functionality of the AI controller; (2) using an Evolution Strategy-based algorithm to synthesize the blank coefficients in the sketch.

\subsubsection{Sketching} Our approach models the synthesis of proxy programs for AI controllers as a sketch-based program synthesis problem~\cite{solar2009sketching}. A sketch is a partial code template with blank holes that are to be filled in. The synthesizer then searches for statements or parameters that can fill these holes. In our study, we create a sketch following~\cite{zhu2019inductive} that allows us to implement one or a few linear controllers over a collection of variables representing the system states of the control system. The sketch is parameterized by a matrix of coefficients $\theta$ and is defined as follows:
\begin{lstlisting}[mathescape]
def Proxy[$\theta$]($x_1$, $x_2$, ..., $x_n$):
    $y_1$ = $\theta_{[1,1]} * x_1$ + $\theta_{[1,2]} * x_2$ + ... + $\theta_{[1,n]} * x_n$ + $\theta_{[1,n+1]}$
    $y_2$ = $\theta_{[2,1]} * x_1$ + $\theta_{[2,2]} * x_2$ + ... + $\theta_{[2,n]} * x_n$ + $\theta_{[2,n+1]}$
    ...
    $y_m$ = $\theta_{[m,1]} * x_1$ + $\theta_{[m,2]} * x_2$ + ... + $\theta_{[m,n]} * x_n$ + $\theta_{[m,n+1]}$
    return $y_1$, $y_2$, ..., $y_m$
\end{lstlisting}
where $x_1, x_2, ..., x_n$ are the system state variables in control systems, $y_1, y_2, ..., y_m$ are the outputs of the synthesized program. $\theta$ is a matrix of coefficients with size $m \times (n+1)$, where $m$ and $n$ are the number of outputs of the AI controller and the number of system state variables in the control system, respectively. $\theta_{[i,j]}$ represents the $j$-th coefficient in the $i$-th equation. Note that the last coefficient $\theta_{[*, n+1]}$ is a constant term that is not associated with any variable. Such a constant can be viewed as a bias term that can allow linear programs to express a wider range of functions~\cite{kwok1996use}. This definition allows us to flexibly create program sketches for any control system by adjusting the number of equations and variables, as long as the system's inputs and outputs are known. We believe this can be done with low difficulty.

Although our sketch implementing linear controllers may resemble a simple neural network with weights and biases, there are key differences in their representations, with ours being simpler and more efficient. Specifically, our linear controller computes the output as a linear combination of the inputs, typically expressed as $y = W \cdot x + b$, where $W$ is the coefficient matrix, $x$ is the input vector, and $b$ is the bias. In contrast, a neural network generally consists of multiple layers, each potentially incorporating nonlinear activation functions (e.g., tanh and sigmoid). Even a basic feedforward neural network with only one input layer and one hidden layer involves more complex computations than a linear controller, represented as $y = \sigma(W_2 \cdot (\sigma(W_1 \cdot x + b_1) + b_2))$, where $W_1$ and $W_2$ are the weight matrices, $b_1$ and $b_2$ are the bias vectors in the input and hidden layers, and $\sigma$ is a nonlinear activation function. Neural networks with this structure often incur higher computational costs during prediction and potentially take more time to synthesize, so we opt for linear controllers in our sketch for greater simplicity and efficiency.

\begin{algorithm}[!t]
    \small
    \caption{Evolution Strategy-based Synthesis}
    \label{algo:sketch-es}
    \SetKwInput{Input}{Require}
    \SetKwInput{Para}{Parameters}
    \SetKwComment{Comment}{/* }{ \texttt{*/}}
    \KwInput{AI-enabled control system $\mathcal{C}$, AI controller $A^c$, Proxy program sketch $\mathcal{P}[\theta](x_1, ..., x_n)$ with blank coefficients $\theta$}
    \Para{Maximum number of time steps $t$, Maximum number of iterations $k$, Population size $m$, Noise standard deviation $\sigma$, and Learning rate $\alpha$}
    \SetKwFunction{FRandomS}{\sc RandomGenerate}
    \SetKwFunction{FConcatenate}{\sc Concatenate}
    \SetKwFunction{FMean}{\sc Mean}
    \SetKwFunction{FStd}{\sc StandardDeviation}
    \SetKwFunction{FDis}{\sc Distance}
    $\theta \gets $ \FRandomS{$|\theta|$} \label{line:random1} \Comment*[r]{\it Initialize coefficients}
    \Repeat{\upshape{\sc Maximum number of iterations $k$ is reached}}{
    $\boldsymbol{N} \gets \FRandomS(m/2, |\theta|)$ \label{line:random2} \Comment*[r]{\it Generate $m\times|\theta|$ random noise by mirror sampling}
    $\boldsymbol{N} \gets \FConcatenate(\boldsymbol{N}, -\boldsymbol{N})$ \label{line:concatenate} \\
    \For(\Comment*[f]{\it Update coefficients for $t$ time steps}){$j \gets 1$ \KwTo $t$}{
        \For{$i \gets 1$ \KwTo $m$}{
        $\theta^\prime \gets \theta + \boldsymbol{N}_{i}$ \label{line:perturb} \Comment*[r]{\it Perturb the current coefficients}
        $\boldsymbol{R} \gets \FDis(\mathcal{C}, A^c, \mathcal{P}[\theta'](x_1, ...,x_n))$ \label{line:fitness} \Comment*[r]{\it Calculate the fitness}
        }
    $\boldsymbol{R} \gets (\boldsymbol{R} - \FMean(\boldsymbol{R}))/ \FStd(\boldsymbol{R})$ \label{line:standardize} \Comment*[r]{\it Standardizes the fitness}
    $\theta \gets \theta + \alpha \frac{1}{m\sigma} \sum_{i=1}^{m} \boldsymbol{N}_{i} \cdot \boldsymbol{R}_{i}$ \label{line:update} \Comment*[r]{\it Update coefficients}
    }
    }
    \Return $\mathcal{P}[\theta](x_1, ...,x_n)$ \Comment*[r]{\it Program with synthesized coefficients}
\end{algorithm}

\subsubsection{Evolution Strategy} Once we have constructed the sketch, we then employ an Evolution Strategy (ES)-based synthesis algorithm to synthesize the coefficients in the sketch. ES is a well-known optimization algorithm that has been successfully applied to solve program synthesis~\cite{sobania2022comprehensive} and controller synthesis problems~\cite{salimans2017evolution}.
The ES-based synthesis algorithm is presented in Algorithm~\ref{algo:sketch-es}. The algorithm first randomly initializes the coefficients $\theta$ in the sketch (line~\ref{line:random1}), then performs the following steps iteratively:
\begin{itemize}
    \item It first generates a matrix of random noise $\boldsymbol{N}$ of size $m \times |\theta|$, from a standard normal distribution (line~\ref{line:random2}). When generating the matrix, we adopt mirror sampling~\cite{brockhoff2010mirrored,mania2018simple} to improve the performance of the algorithm. The algorithm first generates a random noise matrix of size $\frac{m}{2} \times |\theta|$, and then mirrors the matrix by negating each value to get another matrix with the same size. The algorithm then concatenates the two matrices to obtain a matrix of random noise of size $m \times |\theta|$ (line~\ref{line:concatenate});
    \item For each row in the random noise matrix $\boldsymbol{N}$, the algorithm perturbs the current coefficients $\theta$ by adding the row to the current coefficients $\theta$ and obtains a new matrix of coefficients $\theta^\prime$ (line~\ref{line:perturb});
    \item The algorithm proceeds by obtaining the fitness matrix $\boldsymbol{R}$ for each perturbed coefficient $\theta^\prime$. It executes the control system $\mathcal{C}$ using both the proxy program $\mathcal{P}[\theta^\prime](x_1, ...,x_n)$ and the AI controller $A^c$ for once. The fitness is then computed as the distance between the outputs of the proxy program and the AI controller (line~\ref{line:fitness});
    \item The algorithm standardizes the fitness $\boldsymbol{R}$ (line~\ref{line:standardize}) and then updates the coefficients $\theta$ by the following update function (line~\ref{line:update}):
    \begin{equation}
    \theta \leftarrow \theta + \alpha \frac{1}{m\sigma} \sum_{i=1}^{m} \boldsymbol{N}_{i} \cdot \boldsymbol{R}_{i}
    \end{equation}
    where $\alpha$ is the learning rate, $\sigma$ is the noise standard deviation, and $\boldsymbol{N}_{i}$ and $\boldsymbol{R}_{i}$ are the $i$-th row of the random noise matrix $\boldsymbol{N}$ and the fitness $\boldsymbol{R}$, respectively. Such an update is to perform a stochastic gradient estimate step introduced in~\cite{salimans2017evolution}, which has been shown to be effective in optimizing coefficients and achieving better performance than gradient-based optimization algorithms~\cite{salimans2017evolution, mania2018simple}. Unlike gradient-based methods, however, this algorithm does not require gradient information derived from partial derivatives, making it more efficient in certain cases.
   \end{itemize}
The update procedure is executed for $t$ time steps at each iteration out of the $k$ iterations, i.e, $\theta$ is updated $n \times k$ times. The algorithm either terminates when the maximum number of iterations $k$ is reached or the fitness of the proxy program $\mathcal{P}[\theta](x_1, ...,x_n)$, i.e., the distance between the outputs of the proxy program and the AI controller, is smaller than a predefined threshold, which indicates that the synthesized program is sufficiently similar to the AI controller. The algorithm returns the proxy program $\mathcal{P}[\theta](x_1, ...,x_n)$ with the synthesized coefficients $\theta$.
All the hyperparameter settings in this algorithm can be found in Section~\ref{sec:benchmarks}.


Here one may wonder why we do not directly perform formal verification on the synthesized proxy program but instead perform falsification with it. This is because safety specifications are typically built on the plant rather than the controllers. The outputs of the AI controllers or proxy programs do not always reflect the safety properties of the system. Therefore, we apply falsification to analyze the control system to identify safety violations.

\begin{algorithm}[!t]
    \small
    \caption{$\epsilon$-greedy Strategy-based Sub-specification Sampling}
    \label{algo:epsilon-greedy}
    \SetKwComment{Comment}{/* }{ \texttt{*/}}
    \SetKwInput{Input}{Require}
    \SetKwInput{Para}{Parameters}
    \Input{Sub-specifications $\varPhi$, Historical lowest robustness semantics for each sub-specification $\mathcal{R}$}
    \Para{Probability threshold $\epsilon$}
    \SetKwFunction{FRandom}{\sc Random}
    \SetKwFunction{FRandomS}{\sc RandomSample}
    \eIf{$\FRandom(0, 1) < \epsilon$\label{algo:epsilon-greedy:random}}{
        $\varphi \gets \FRandomS(\varPhi)$ \label{algo:epsilon-greedy:random-sample} \Comment*[r]{\it Sample one sub-specification randomly}
    }
    {
        $i \gets \arg\min_{i\in \{1,...|\mathcal{R}|\}} \mathcal{R}$ \label{algo:epsilon-greedy:exploit} \Comment*[r]{\it Sample the sub-specification with the lowest robustness}
        $\varphi \gets \varPhi_i$ \label{algo:epsilon-greedy:exploit-sample} \\
    }
    \Return $\varphi$ \Comment*[r]{\it Return a sub-specification for subsequent falsification}
\end{algorithm}

\subsection{Sampling}
\label{subsec:sampling}

To achieve more comprehensive coverage of the conjunctive safety specification during falsification, \tool first separates the conjunctive specifications into sub-specifications. We can then select one sub-specification from the set of all sub-specifications to falsify. However, blindly exploring the sub-specification that has not been falsified to falsify is not effective, as some sub-specifications are more difficult to falsify than others or even impossible to falsify. Always forcing the falsification process to work on such sub-specifications can result in a waste of resources. Conversely, always exploiting previously falsified sub-specifications may not be effective either, as it may result in low sub-specification violation coverage. Thus, we propose an $\epsilon$-greedy Strategy-based Sampling method to intelligently decide which sub-specifications to falsify, balancing exploration and exploitation, thereby boosting the sub-specification coverage effectively.

The basic concept of the $\epsilon$-greedy strategy has been introduced in Section~\ref{subsec:epsilon-greedy}. Here, we describe its application in the context of falsification, as outlined in Algorithm~\ref{algo:epsilon-greedy}. The primary adaptations of the $\epsilon$-greedy strategy to the sub-specification sampling process involve two key changes: (1) defining the reward as the historical robustness semantics of each sub-specification from prior falsification attempts--rather than using expected rewards, we track the lowest robustness value for each sub-specification; and (2) modifying the exploitation phase to select the sub-specification with the lowest robustness value instead of the highest reward, since a lower robustness value indicates a higher likelihood that the sub-specification may be violated in future attempts.

Given a set of sub-specifications $\varPhi$, our algorithm begins by determining whether to explore by randomly selecting a sub-specification, using a randomly-generated number  (line~\ref{algo:epsilon-greedy:random}). If this number is less than the exploration probability $\epsilon$, the algorithm randomly selects a sub-specification from $\varPhi$ (line~\ref{algo:epsilon-greedy:random-sample}). Otherwise, it proceeds with exploitation, selecting the sub-specification with the lowest robustness value based on the historical robustness semantics $\mathcal{R}$ (line~\ref{algo:epsilon-greedy:exploit}). The selected sub-specification is then returned for subsequent falsification. The algorithm is called at each iteration of the falsification process to decide which sub-specification to falsify.



\subsection{Falsification \& Refinement}
\label{subsec:refinement}

\subsubsection{Falsification}
We employ the Simulated Annealing (SA)-based falsification algorithm in \tool, which is already described in Algorithm~\ref{algo:SA}.
Note that in \tool the falsification algorithm is performed on the control system with a synthesized proxy program rather than the AI controller. Since the proxy program is not with the exact same functionality as the original AI controller, the SA-based falsification algorithm may find some spurious violations. Therefore, we use the original AI controller to check whether the violations are spurious or not. If the violations are spurious, we refine the proxy program to perform more similarly to the original AI controller. Otherwise, we terminate the falsification process and report the found violations.

\begin{algorithm}[!t]
    \small
    \caption{Refinement of Synthesized Proxy Program}
    \label{algo:refinement}
    \SetKwInput{Input}{Require}
    \SetKwInput{Para}{Parameters}
    \SetKwComment{Comment}{/* }{ \texttt{*/}}
    \KwInput{AI-enabled control system $\mathcal{C}$, AI controller $A^c$, Proxy program sketch $\mathcal{P}[\theta](x_1, ..., x_n)$ with synthesized coefficients $\theta$, Spurious falsifying input $s$}
    \Para{Maximum number of time steps $t$, Population size $m$, Noise standard deviation $\sigma$, and Learning rate $\alpha$}
    \SetKwFunction{FConcatenate}{\sc Concatenate}
    \SetKwFunction{FMean}{\sc Mean}
    \SetKwFunction{FStd}{\sc StandardDeviation}
    \SetKwFunction{FDis}{\sc Distance}
    $\boldsymbol{N} \gets \FRandomS(m/2, |\theta|)$ \label{line:random2-refine} \Comment*[r]{\it Generate $m\times|\theta|$ random noise by mirror sampling}
    $\boldsymbol{N} \gets \FConcatenate(\boldsymbol{N}, -\boldsymbol{N})$ \label{line:concatenate-refine} \\
    $\mathcal{C} \gets \mathcal{C}(s)$ \label{line:control-system-refine} \Comment*[r]{\it Initialize the control system with the spurious falsifying input}
    \For(\Comment*[f]{\it Update coefficients for $t$ time steps}){$j \gets 1$ \KwTo $t$}{
        \For{$i \gets 1$ \KwTo $m$}{
        $\theta^\prime \gets \theta + \boldsymbol{N}_{i}$ \label{line:perturb-refine} \Comment*[r]{\it Perturb the current coefficients}
        $\boldsymbol{R} \gets \FDis(\mathcal{C}, A^c, \mathcal{P}[\theta'](x_1, ...,x_n))$ \label{line:fitness-refine} \Comment*[r]{\it Calculate the fitness}
        }
    $\boldsymbol{R} \gets (\boldsymbol{R} - \FMean(\boldsymbol{R}))/ \FStd(\boldsymbol{R})$ \label{line:standardize-refine} \Comment*[r]{\it Standardizes the fitness}
    $\theta \gets \theta + \alpha \frac{1}{m\sigma} \sum_{i=1}^{m} \boldsymbol{N}_{i} \cdot \boldsymbol{R}_{i}$ \label{line:update-refine} \Comment*[r]{\it Update coefficients}
    }
    \Return $\mathcal{P}[\theta](x_1, ...,x_n)$ \Comment*[r]{\it Program with refined coefficients}
\end{algorithm}

\subsubsection{Refinement} Once we have a spurious violation, we use the spurious falsifying input to refine the proxy program, i.e., slightly changing the coefficients in the proxy program to make its outputs more similar to those of the AI controller. The refinement procedure follows Algorithm~\ref{algo:refinement}. It still follows the ES-based synthesis presented in Algorithm~\ref{algo:sketch-es}, but introduces two major differences. First, instead of starting with a random matrix, the refinement begins with the current proxy program and its synthesized coefficients. Second, rather than initializing the control system randomly and collecting system states for fitness evaluation, the spurious falsifying input $s$ is used to initialize the system, after which fitness evaluation occurs (lines~\ref{line:control-system-refine}-\ref{line:fitness-refine}). The rest of the procedure remains the same as the ES-based synthesis procedure. Importantly, the refinement is applied only once after a spurious violation is detected during the falsification process to avoid overfitting the proxy program to the spurious input.
Such a refinement procedure is inspired, at a high level, by the counterexample-guided abstraction refinement loop in formal verification~\cite{clarke2000counterexample,menghi2020approximation,shariffdeen2021concolic}. The difference is that instead of using the counterexample to refine a model abstraction, we use the spurious falsifying input to refine the synthesized proxy program.

%% file: sections/experiments.tex
\section{Evaluation Settings}
\label{sec:exp_setup}

Our evaluation aims to answer the following research questions:

\vspace{0.3em}
\noindent\textbf{RQ1 (Effectiveness): How effective is \toolbf in revealing violations of safety specifications?}
Although \tool is highlighted to resolve the issues of efficiency and sub-specification coverage in falsification of AI-enabled control systems, it is still unclear how effective \tool is in finding safety violations.
We evaluate the ability of \tool in 8 AI-enabled control systems and compare it to our baseline \baseline~\cite{thibeault2021psy} which is introduced in Section~\ref{sec:benchmarks}.

\vspace{0.3em}
\noindent\textbf{RQ2 (Efficiency): How efficient is \toolbf in detecting violations of safety specifications?}
One of the most innovative aspects of \tool is its use of a synthesized proxy program to address the scalability issue in falsification of AI-enabled control systems. To demonstrate the advantage of this approach, we evaluate the time efficiency of \tool by comparing it to the baseline \baseline that does not incorporate program synthesis.

\vspace{0.3em}
\noindent\textbf{RQ3 (Sub-specification Violation Coverage): How comprehensive is \toolbf in covering safety specifications?}
Another key feature of \tool is its capability to cover many sub-specifications within conjunctive specifications. To demonstrate this advantage, we compare the sub-specification violation coverage of \tool (i.e., how many sub-specifications in conjunctive safety specifications can be covered by its detected violations) with that of the baseline that does not employ sub-specification sampling.

\subsection{Benchmark \& Baseline}
\label{sec:benchmarks}

\begin{table}[t!]
    \caption{Control systems and their safety specifications. The detailed description of these systems and specifications are in Section~\ref{sec:benchmarks}. The table also shows the number of sub-specifications, inputs, outputs, and neural networks sizes of AI controllers.}
    \label{tab:benchmarks}
    \footnotesize
    \centering
    \renewcommand{\familydefault}{\sfdefault}\normalfont
    \scalebox{0.92}{
    \begin{tabular}{c|c|c|c|c|c}
    \hline
    \hline
    \multirow{2}{*}{\begin{tabular}[c]{@{}c@{}}Control\\ System\end{tabular}} & \multirow{2}{*}{Safety Specification} & \multirow{2}{*}{\begin{tabular}[c]{@{}c@{}}\#Sub- \ \ \\Specifications\end{tabular}} & \multirow{2}{*}{\begin{tabular}[c]{@{}c@{}}\#Inputs/\\ \#Outputs\end{tabular}} & \multicolumn{2}{c}{AI Controller Size} \\
    \cline{5-6}
     &  &  &  & \#Layers & \#Neurons \\ \hline\hline
    CartPole   & \makecell{$\square_{[0, 200]} ( \vert\eta\vert <90^{\circ} \wedge \vert\delta \vert <0.3 \wedge$ \\ $\vert v_1 \vert < 0.3 \wedge \vert v_2 \vert < 0.5 )$} & 8& 4/1 & 3 & $[300, 250, 200]$ \\ \hline
    Pendulum    & $\square_{[0, 200]} \left( \vert\eta\vert <90^{\circ} \wedge\vert \omega \vert <\pi/2 \right)$ & 4 & 2/1 & 3 & $[280, 240, 200]$ \\ \hline
    Quadcopter  & $\square_{[0, 300]} \left( \vert\eta_1\vert <\pi/2 \wedge\vert \eta_2 \vert <\pi/2 \right)$  & 4 & 2/1 & 2 & $[240, 200]$ \\ \hline
    Self Driving & $\square_{[0, 200]} \left( \vert\eta\vert <90^{\circ} \wedge\vert d\vert < 2.0 \right)$   & 4 & 2/1 & 3& $[300, 250, 200]$   \\ \hline
    Lane Keeping & $\square_{[0, 300]} \left(\vert d \vert < 0.9\right)$  & 2& 4/2 & 2& $[240, 200]$ \\ \hline
    4-Car Platoon & \makecell{$\square_{[0, 200]} (\vert a \vert <= 2.0 \wedge \vert b \vert <= 0.5 \wedge \vert c \vert <= 0.35 \ \wedge$ \\ $\vert d \vert <= 0.5 \wedge \vert e \vert <= 1.0 \wedge \vert f \vert <= 0.5 \wedge \vert g \vert <= 1.0)$} & 14 & 7/4 & 3& $[500, 400, 300]$ \\ \hline
    8-Car Platoon & \makecell{$\square_{[0, 200]} (\vert a \vert <= 2.0 \wedge \vert b \vert <= 0.5 \wedge \vert c \vert <= 1.0 \ \wedge$ \\ $\vert d \vert <= 0.5 \wedge \vert e \vert <= 1.0 \wedge \vert f \vert <= 0.5 \wedge \vert g \vert <= 1.0 \ \wedge$ \\ $\vert h \vert <= 0.5 \wedge \vert i \vert <= 1.0 \wedge \vert j \vert <= 0.5 \wedge \vert k \vert <= 1.0 \ \wedge$ \\ $\vert l \vert <= 0.5 \wedge \vert m \vert <= 1.0 \wedge \vert n \vert <= 0.5 \wedge \vert o \vert <= 1.0)$} & 30 &15/8 & 3 & $[400, 300, 200]$  \\ \hline
    Oscillator & $\square_{[0, 300]} \left(a < 0.05 \right)$ & 1 & 18/2 & 3 & $[280, 240, 200]$  \\
    \hline \hline
    \end{tabular}
    }
\end{table}

\subsubsection{Benchmark}
Table~\ref{tab:benchmarks} shows 8 of the 15 control systems from Zhu et al.~\cite{zhu2019inductive}. We adopt these systems in our evaluation and exclude the other 7 as they already show perfect performance, i.e., never violate any safe specification, according to the experiments in~\cite{zhu2019inductive}.
The 8 selected ones are representative of real-world control systems, such as Segway transporters~\cite{Segway}, camera drones~\cite{DJI}, and autonomous driving cars~\cite{Udacity}, and all are end-to-end systems that feature one single AI controller using raw input signals to generate control signals. We introduce the 8 control systems as follows:
\begin{itemize}
    \item {\sf CartPole}: This system consists of a pole attached to an unactuated joint connected to a cart moving along a track. Safety is ensured if the pole's angle $\eta$ is less than 90$^\circ$ from the upright position, the cart distance $\delta$ from the origin is within 0.3 meters, and the velocities $v_1$ and $v_2$ are below 0.3 and 0.5 m/s, respectively.
    \item {\sf Pendulum}: This system is to maintain a pendulum attached to an unactuated joint in an upright position. Safety is ensured if the pendulum's angle $\eta$ does not exceed the threshold of 90$^\circ$ or the pendulum's angular velocity $\omega$ remains below $\pi/2$ rad/s.
    \item {\sf Quadcopter}: The control system aims to achieve stable flight for a quadcopter by limiting the angle $\eta_1$ of the pitch rotation and the angle $\eta_2$ of the yaw rotation not to exceed $\pi/2$.
    \item {\sf Self Driving}: Discussed in detail in Section~\ref{sec:motivation}.
    \item {\sf Lane Keeping}: This system aims to keep a vehicle between lane markers and centered in a possibly curved lane. Safety is ensured if a safe distance $d$ to the cars in front is less than 0.9 meters.
    \item {\sf 4-Car Platoon} and {\sf 8-Car Platoon}: These two systems model 4 or 8 vehicles forming a platoon and maintaining a safe relative distance between them. Each variable in the specification is the distance between two consecutive vehicles in the platoon, and safety is assured if they are below certain thresholds.
    \item {\sf Oscillator}: This system consists of a two-dimensional switched oscillator and a 16-order filter. Safety is achieved if the output signal $a$ of the filter remains below the safe threshold of 0.05.
\end{itemize}
Note that the inputs for these systems adhere to specific predefined ranges. The training data of AI controllers as well as the falsifying inputs are both from those ranges. The other falsifying inputs, which are outside the input ranges, are not considered within the scope of our work.
Moreover, these control systems exhibit significant variation in the number of sub-specifications, ranging from 1 to 30 as shown in Table~\ref{tab:benchmarks}. This diversity allows for a comprehensive evaluation. Detailed information regarding the input ranges and sub-specifications for each control system can be found in our replication package, accessible through the link provided in Section~\ref{sec:conclusion}.

We utilized the well-trained AI controllers for the 8 control systems from the official implementation of Zhu et al.~\cite{zhu2019inductive}, where these controllers were trained by Deep Deterministic Policy Gradient~\cite{LillicrapHPHETS15}, a popular DRL algorithm for continuous control tasks.
Table~\ref{tab:benchmarks} shows the numbers of input and output variables of the AI controllers as well as their neural network sizes. The AI controllers have a wide range of sizes, featuring 2 or 3 layers with neuron counts varying from 440 (240+200) to 1200 (500+300+400); such sizes are sufficiently large for AI controllers (e.g., those controllers used in~\cite{zhang2022falsifai, tran2020nnv} have no more than 400 neurons).

\subsubsection{Baseline}
We compare \tool to \baseline~\cite{thibeault2021psy}, a state-of-the-art and open-source tool for falsification of control systems. \baseline recently won the 2022 ARCH-COMP prize for its technical achievements and excellent performance in falsification~\cite{ernst2022arch}. It is the Python version of a MATLAB tool called S-TaLiRo~\cite{annpureddy2011s}, which has been classified as the most mature tool ready for industrial deployment~\cite{kapinski2016simulation,tuncali2018experience}. In our study, we use this Python version of \baseline to ensure consistency with the programming language of the Python-developed control systems we selected. \baseline provides built-in optimization algorithms, including Uniform Random and Simulated Annealing. We chose the Simulated Annealing-based algorithm to support \baseline in our experiments, as it has been shown to be more effective than Uniform Random~\cite{ernst2022arch,song2022cyber}.


In our evaluation, we did not include several other open-source black-box falsification tools, such as Breach~\cite{donze2010breach}, ARIsTEO~\cite{menghi2020approximation}, FalStar~\cite{ernst2021falsification}, ForeSee~\cite{zhang2021effective}, zlscheck~\cite{ismailbennani}, NNV~\cite{tran2020nnv}, and Mosaic~\cite{xie2023mosaic}. These tools are specialized for MATLAB/Simulink models and written in different programming languages such as MATLAB, Java, Scala, or OCaml. The features of different programming languages can greatly impact the experimental results. Also, integrating these tools into our evaluation could pose significant compatibility challenges with the Python-developed control systems and toolchains used in our experiments.
We believe that comparing \tool to \baseline, the only existing falsification toolbox natively in Python, is fair and appropriate since \baseline is a state-of-the-art and well-established falsification tool that has been well studied.
The 2022 ARCH-COMP prize awarded to \baseline also demonstrates its advancement and maturity in falsification~\cite{ernst2022arch}.

\subsection{Implementation and Hyperparameters}
\label{sec:implementation}

\tool is built on top of the implementation of Simulated Annealing (SA)-based falsification algorithm and other infrastructure components from \baseline (version 1.0.0b3).
To compute the quantitative robustness semantics, we integrate an off-the-shelf tool called RTAMT~\cite{nivckovic2020rtamt} (version 0.3.0) into our framework.
We set the maximum number of iterations for the SA-based algorithm to 100, a recommended value by recent literature~\cite{menghi2020approximation,yamagata2020falsification}. By reusing all other default hyperparameters of \baseline in the SA-based algorithm, we ensure a fair comparison while also evaluating the unique design of \tool.
For the Evolution Strategy (ES)-based synthesis algorithm, we use a population size of 50, and both the maximum number of iterations and time steps are set to 100. The noise standard deviation and the learning rate are set to 0.1 and 0.05, respectively. The threshold of the fitness is set to 0.1. The refinement follows the same hyperparameters as the ES-based synthesis algorithm, except that the learning rate is set to 1e-3 to avoid overfitting.
In our $\epsilon$-greedy sampling strategy, we set the probability threshold $\epsilon$ to 0.9. All the experiments are conducted using an Ubuntu 18.04 server equipped with an Intel Xeon E5-2698 CPU and 504 GB RAM.

\subsection{Experimental Setup and Metrics}
\label{sec:metrics}

For a comprehensive evaluation, we employ two settings of the budgets of falsification in our experiments, as follows:
\begin{itemize}
    \item {\bf Setting 1:} Following the common practice of the annual ARCH workshop~\cite{ernst2020arch,ernst2021arch}, we specify the budget as the maximum number of falsification trials, which is set to 50.
    \item {\bf Setting 2:} Following the common practice of search-based testing~\cite{kang2022test,pang2022mdpfuzz,zolfagharian2023search}, we specify the budget as a fixed duration of time, which is set to 10 minutes, in line with prior studies~\cite{kang2022test,soltani2018search}.
\end{itemize}
To mitigate the impact of randomness, we repeat each experiment 10 times and report the average results along with their statistical significance and effect size as suggested by Arcuri and Briand~\cite{arcuri2011practical}.

Then for each setting, we compare the performance of \tool and \baseline using the following metrics:
\begin{itemize}
    \item {\bf Setting 1:} The effectiveness of each tool is evaluated by the falsification success rate, which is the ratio of the number of trials that uncover violations to 50 trials. This metric is commonly adopted in falsification studies~\cite{ernst2022arch,ernst2021arch,zhang2022falsifai}. The efficiency of each tool is measured by the falsification time, which is the time required to complete 50 falsification trials.
    \item {\bf Setting 2:} The effectiveness of each tool is evaluated using the metric of the number of violations found by each tool within the given time budget of 10 minutes, following prior search-based testing studies~\cite{sun2022lawbreaker,kang2022test}. The efficiency of each tool is measured by the time required to reveal one safety violation.
\end{itemize}
Additionally, we also evaluate the sub-specification violation coverage of each tool, which is already defined in Section~\ref{sec:motivation}. Note that we report the performance improvements of \tool over \baseline in terms of the relative percentage difference in metrics. This is calculated by taking the difference between the two tools' performance and dividing it by the performance of \baseline.

%% file: sections/results.tex
\section{Evaluation Results}
\label{sec:results}


\subsection{RQ1: Effectiveness of {\sc \textbf{Synthify}}}

\begin{table*}[t!]
    \footnotesize
    \centering
    \caption{Falsification results of {\sf \textsc{Synthify}} and {\sf \textsc{PSY-TaLiRo}} on the 8 control systems.
    The $\hat{A}_{12}$ column shows the effect size, where $N$, $M$, and $L$ indicate negligible, medium, and large effect sizes, respectively.}
    \label{tab:RQ1}
    \renewcommand{\familydefault}{\sfdefault}\normalfont
    \scalebox{0.89}{
    \begin{tabular}{@{}c|c|c|c|c|c|c|c|c@{}}
    \hline
    \hline
    & \multicolumn{4}{c|}{Setting 1 (50 trials)}  & \multicolumn{4}{c}{Setting 2 (10 minutes)}  \\ \hline\hline
    System & \baseline & \tool           & {\it p}-value     & {\it \^A}\textsubscript{12} & \baseline & \tool                  & {\it p}-value     & {\it \^A}\textsubscript{12} \\ \hline
    CartPole    & 99.8\%            & \textbf{100.0\% (+0.2\%)}  & 0.36  & {\it N} (0.55)     & 99.8             & \textbf{561.1 (5.6$\times$)}  & 1.82e-4\textless{}0.01 & {\it L} (1.00)     \\ \hline
    Pendulum     & 0.0\%             & \textbf{100.0\% (---)} & 1.59e-5\textless{}0.01 & {\it L} (1.00)     & 0.0               & \textbf{23.9 (---)}                 & 4.84e-5\textless{}0.01 & {\it L} (1.00)     \\ \hline
    Quadcopter  & 99.0\%            & \textbf{100.0\% (+1.0\%)}  & 0.07                   & {\it M} (0.65)     & 82.4             & \textbf{177.1 (2.1$\times$)}  & 5.20e-4\textless{}0.01 & {\it L} (0.98)     \\ \hline
    Self-Driving & 96.2\%            & \textbf{100.0\% (+4.0\%)}  & 7.25e-4\textless{}0.01 & {\it L} (0.90)     & 59.9             & \textbf{150.9 (2.5$\times$)} & 1.86e-4\textless{}0.01 & {\it L} (0.80)     \\ \hline
    Lane Keeping & 0.0\%            & \textbf{100.0\% (---)}  & 1.59e-5\textless{}0.01 & {\it L} (1.00)     & 0.0             & \textbf{159.3 (---)} & 6.20e-5\textless{}0.01 & {\it L} (1.00)     \\ \hline
    4-Car Platoon & 42.0\%            & \textbf{100.0\% (+138.1\%)}  & 6.25e-5\textless{}0.01 & {\it L} (1.00)     & 5.7             & \textbf{626.6 (109.9$\times$)} & 1.82e-4\textless{}0.01 & {\it L} (1.00)     \\ \hline
    8-Car Platoon & 0.0\%            & \textbf{100.0\% (---)}  & 1.59e-5\textless{}0.01 & {\it L} (1.00)     & 0.0             & \textbf{472.0 (---)} & 6.39e-5\textless{}0.01 & {\it L} (1.00)     \\ \hline
    Oscillator & 99.0\%            & \textbf{100.0\% (+1.0\%)}  & 1.37e-2\textless{}0.05 & {\it L} (0.75)     & 90.1             & \textbf{470.7 (5.2$\times$)} & 1.82e-4\textless{}0.01 & {\it L} (1.00)     \\
    \hline \hline
    Average      & 54.5\%            & \textbf{100.0\% (+83.5\%)} & ---                    & ---            & 42.2             & \textbf{330.2 (7.8$\times$)}   & ---                    & ---            \\ \hline\hline
    \end{tabular}
    }
\end{table*}

Table~\ref{tab:RQ1} shows the falsification results of \tool and \baseline for the 8 control systems.
As mentioned in Section~\ref{sec:exp_setup}, we repeat each experiment 10 times and report the average results here.
For Setting 1, we compare the falsification success rates of the two tools and find that \tool outperforms \baseline on all control systems. \tool achieves a 100\% falsification success rate on all control systems. In contrast, \baseline only achieves an average success rate of 54.5\%. On average, \tool outperforms \baseline by 83.5\% ((100\% - 54.5\%)/54.5\% = 83.5\%) in terms of the falsification success rate.

To account for the impact of randomness and ensure the credibility of the results, we used the Mann-Whitney U test~\cite{mann1947test}, which is a non-parametric statistical test used to compare two independent groups. We use it to measure the statistical significance of the difference in falsification success rates between the two tools across the 10 experiments. Our results show that \tool significantly outperforms \baseline on 6 out of 8 control systems, with a confidence level of 99\% or 95\%. On {\sf CartPole} and {\sf Quadcopter}, the improvements achieved by \tool are not as significant because the falsification success rates of \baseline are already quite high.

We also use the Vargha-Delaney statistic $\hat{A}_{12}$~\cite{vargha2000critique} to measure the effect size. This non-parametric measure gives the probability that a randomly chosen value from one group is higher or lower than one from another group. It is commonly used to evaluate search-based testing methods due to their randomness~\cite{palomba2016automatic,meng2022linear,olsthoorn2020generating}. According to Vargha and Delaney~\cite{vargha2000critique}, $\hat{A}_{12}$ values greater than 0.71 (or less than 0.29) indicate a ``large'' effect size. Values between 0.64 and 0.71 (or between 0.29 and 0.36) correspond to a ``medium'' effect size, while values between 0.36 and 0.64 suggest a ``negligible'' effect size. As shown in Table~\ref{tab:RQ1}, \tool significantly outperforms \baseline with a large effect size on 6 out of 8 control systems. Only on {\sf CartPole} and {\sf Quadcopter}, we observe negligible and medium effect sizes. But overall, the results show that \tool is more effective than \baseline in finding violations of safety specifications.

In Setting 2, \tool also consistently outperforms \baseline across all control systems. Table~\ref{tab:RQ1} presents the results of Setting 2, where \tool and \baseline are compared in terms of the number of violations they reveal within the same time budget of 10 minutes. The results show that, on average, \tool can uncover 330.2 violations of safety specifications within 10 minutes, which is 7.8 times more than \baseline (42.3 violations).
This result highlights \tool's ability to reveal significantly more safety violations than \baseline within the same time budget. Furthermore, the results of Mann-Whitney U tests and $\hat{A}_{12}$ statistics show that \tool significantly outperforms \baseline with a 99\% confidence level and a large effect size.

\ans{\textbf{Answers to RQ1}: \tool significantly outperforms \baseline in terms of effectiveness. Within the same falsification trials, \tool outperforms \baseline by 83.5\% in terms of falsification success rate. Within the same time budget, \tool reveals 7.8$\times$ more safety violations than \baseline.}

\subsection{RQ2: Efficiency of {\sc \textbf{Synthify}}}

\begin{table*}[t!]
    \footnotesize
    \centering
    \caption{Falsification time of {\sf \textsc{Synthify}} and {\sf \textsc{PSY-TaLiRo}}. The $\hat{A}_{12}$ column shows the effect size, $L$ indicates a large effect size.}
    \label{tab:RQ2}
    \renewcommand{\familydefault}{\sfdefault}\normalfont
    \scalebox{0.93}{
    \begin{tabular}{@{}c|c|c|c|c|c|c|c|c@{}}
        \hline\hline
    & \multicolumn{4}{c|}{Total falsification time over 50 trials}   & \multicolumn{4}{c}{Time required to reveal one violation}  \\ \hline\hline
    System   & \baseline & \tool  & {\it p}-value     & {\it \^A}\textsubscript{12} & \baseline & \tool  & {\it p}-value     & {\it \^A}\textsubscript{12} \\ \hline
    {\sf CartPole}     & 286.5s            & \textbf{88.2s (3.2$\times$)} & 1.82e-4\textless{}0.01 & {\it L} (1.00)              & 6.1s              & \textbf{1.1s (5.5$\times$)}  & 2.83e-3\textless{}0.01 & {\it L} (1.00)              \\ \hline
    {\sf Quadcopter}   & 127.9s             & \textbf{46.6s (2.7$\times$)} & 1.83e-4\textless{}0.01 & {\it L} (1.00)              & 8.3s              & \textbf{4.1s (2.0$\times$)}  & 2.80e-4\textless{}0.01 & {\it L} (1.00)              \\ \hline
    {\sf Self-Driving} & 543.4s            & \textbf{182.8s (3.0$\times$)} & 1.83e-4\textless{}0.01 & {\it L} (1.00)              & 15.4s              & \textbf{5.5s (2.8$\times$)} & 8.62e-4\textless{}0.01 & $L$ (0.97)              \\ \hline
    4-Car Platoon & 1462.7s & \textbf{80.8s (18.1$\times$)} &  1.83e-4\textless{}0.01 & $L$ (1.00)  & 129.2s & \textbf{1.0s (129.2$\times$)} & 3.30e-4\textless{}0.01 & $L$ (0.98)    \\ \hline
    Oscillator & 295.1s  & \textbf{86.5s (3.4$\times$)} &1.83e-4\textless{}0.01 & $L$ (1.00)  &  6.8s & \textbf{1.3s (5.2$\times$)} & 1.31e-3\textless{}0.01 & $L$ (0.93)  \\ \hline \hline
    Average      & 543.1s            & \textbf{97.0s (5.6$\times$)} & ---                    & ---                     & 33.2s              & \textbf{2.6s (12.8$\times$)}  & ---                    & ---                     \\ \hline\hline
    \end{tabular}
    }
\end{table*}

The comparison results of the falsification time over the 50 trials of Setting 1 are shown in Table~\ref{tab:RQ2}. Note that the falsification time of \tool also includes the time for the synthesis and refinement process.
{\sf Pendulum}, {\sf Lane Keeping}, and {\sf 8-Car Platoon} are excluded from this comparison as \baseline failed to falsify their safety specifications in our RQ1 experiments. For the remaining five control systems, we observe that \tool consistently outperforms \baseline in terms of falsification time for all of them.
On average, \tool is 5.6$\times$ faster than \baseline in completing 50 falsification trials. These performance differences are statistically significant at the 99\% confidence level, as confirmed by the Mann-Whitney U tests. Additionally, the $\hat{A}_{12}$ statistics indicate that the effect sizes of the improvements are large in all cases. Thus, \tool offers significant efficiency gains in the falsification process across different control systems.

In Setting 2, \tool also takes much less time to find a single violation compared to \baseline. Table \ref{tab:RQ2} shows the time required for each tool to find a violation, still excluding a few control systems due to that  \baseline' cannot perform any falsification on them. On average, \tool takes only 2.6 seconds to find one single violation, which is 12.8$\times$ faster than \baseline in finding one single violation. Still, the statistical differences between the time taken by the two tools indicate that \tool significantly outperforms \baseline across different control systems at the 99\% confidence level. The $\hat{A}_{12}$ statistics also show that \tool significantly outperforms \baseline with a large effect size.

\ans{\textbf{Answers to RQ2}: \tool significantly outperforms \baseline in terms of falsification time and the time required to find a violation. Specifically, \tool is 5.6$\times$ faster than the baseline at completing 50 falsification trials. Moreover, \tool can find a violation 12.8$\times$ faster than \baseline.}

\begin{table}[t!]
    \small
    \centering
    \caption{
        Sub-specification violation coverage of {\sf \textsc{Synthify}} and {\sf \textsc{PSY-TaLiRo}}. The $\hat{A}_{12}$ column shows the effect size where $N$ and $L$ indicate negligible and large effect size, respectively.}
    \label{tab:RQ3}
    \renewcommand{\familydefault}{\sfdefault}\normalfont
    \begin{tabular}{c|c|c|c|c}
        \hline\hline
    Control System   & \baseline & \tool & {\it p}-value     & {\it \^A}\textsubscript{12} \\ \hline \hline
    {\sf CartPole}     & 50.0\%            & \textbf{98.6\% (+97.2\%)}  & 2.42e-5\textless{}0.01 & {\it L} (1.00)     \\ \hline
    {\sf Pendulum}    & 0.0\%             & \textbf{25.0\% (---)}   & 1.59e-5\textless{}0.01 & {\it L} (1.00)     \\ \hline
    {\sf Quadcopter}   & 75.0\%            & \textbf{100.0\% (+33.3\%)} & 1.59e-5\textless{}0.01 & {\it L} (1.00)     \\ \hline
    {\sf Self-Driving} & 50.0\%            & \textbf{100.0\% (+100.0\%)} & 9.67e-3\textless{}0.01 & {\it L} (0.80)     \\ \hline
    Lane Keeping & 0.0\% & \textbf{90.0\% (---)} & 3.29e-5\textless{}0.01 & {\it L} (1.00) \\ \hline
    4-Car Platoon & 7.1\%  & \textbf{76.4\% (+976.1\%)}  & 4.14e-5\textless{}0.01 & {\it L} (1.00) \\ \hline
    8-Car Platoon & 0.0\% & \textbf{81.0\% (---)} &  4.17e-5\textless{}0.01  & {\it L} (1.00) \\ \hline
    Oscillator & 100\% & 100\%  & 1.00  & {\it N} (0.50) \\ \hline \hline
    Average      & 35.3\%            & \textbf{83.9\% (+137.7\%)}  & ---                    & ---            \\ \hline\hline
    \end{tabular}
\end{table}

\subsection{RQ3: Sub-specification Violation Coverage}

\tool achieves an average sub-specification violation coverage of 83.9\%, which is a substantial improvement of 137.7\% over \baseline's coverage of 35.3\%, as listed in Table~\ref{tab:RQ3}.
Note that the coverage results are consistent for both Setting 1 and Setting 2.
Of all the control systems, only the results on {\sf Oscillator} show no improvement. The reason for this is that {\sf Oscillator} contains only one sub-specification, and both \tool and \baseline can effectively falsify it. However, for all other control systems, \tool outperforms \baseline by at least 33.3\% in terms of sub-specification violation coverage.

Similar to RQ1 and RQ2, we perform Mann-Whitney U tests and $\hat{A}_{12}$ statistics, and the results confirm that \tool significantly outperforms \baseline on all control systems except {\sf Oscillator}, with a 99\% confidence level and a large effect size. These results further highlight the superiority of \tool in achieving higher sub-specification violation coverage than \baseline.

\ans{\textbf{Answers to RQ3}: \tool significantly outperforms \baseline in terms of sub-specification violation coverage. Specifically, \tool can falsify 83.9\% of safety specifications on average, which is a 137.7\% improvement over \baseline.}

%% file: sections/discussion.tex
\section{Additional Analysis}
\label{sec:discussion}

Aside from the main results presented in Section~\ref{sec:results}, we provide additional analyses to further understand the performance of \tool. These analyses include an ablation study to assess the importance of each core module in \tool, a breakdown of the running time for each process in \tool, an evaluation of the alignment between the proxy programs and AI controllers, and an investigation of \tool's performance on control systems with large AI controllers. We also detail the selection for the hyperparameters of our sub-specification sampling strategy, offer a discussion on the practical implications of our work in the era of end-to-end AI-enabled control systems, as well as the limitations of our study.

\subsection{Ablation Study}
We perform an ablation study by comparing the performance of \tool with different variants. Table~\ref{tab:ablation} shows the average results over 50 falsification trials on the 8 control systems. The variants considered in the study are as follows:

\vspace{0.2em}
\noindent\textbf{w/o Refinement.} \tool without the refinement process (described in Section~\ref{subsec:refinement}). Without this step, the synthesized proxy program may not behave sufficiently similarly to the AI controller, impacting the effectiveness of \tool. The success rate of \tool in finding safety violations within 50 falsification trials drops from 100.0\% to 87.5\%, and the sub-specification violation coverage is also reduced by 12.5\% (from 83.9\% to 71.4\%). These results confirm that the refinement process improves the effectiveness of \tool.

\begin{table}[t!]
    \centering
    \small
    \caption{Ablation study of core modules in {\sf \textsc{Synthify}}. Results are averaged over 50 falsification trials on 8 control systems.}
    \label{tab:ablation}
    \renewcommand{\familydefault}{\sfdefault}\normalfont
    \begin{tabular}{@{}c|c|c|c@{}}
        \hline\hline
         & Success Rate & Time & Coverage \\
        \hline\hline
        \tool & 100.0\% & 97.0s & 83.9\% \\
        \hline
        w/o Refinement & 87.5\% & 94.5s & 71.4\% \\
        \hline
        w/o Synthesis \& Refinement & 51.7\% & 781.3s & 53.1\% \\
        \hline
        w/o $\epsilon$-greedy Sampling & 100.0\% & 78.1s & 46.0\% \\
        \hline\hline
    \end{tabular}
\end{table}

\vspace{0.2em}
\noindent\textbf{w/o Synthesis \& Refinement.} \tool without the synthesis (described in Section~\ref{subsec:synthesis}) and refinement (Section~\ref{subsec:refinement}) processes. These two steps are designed to improve the scalability of falsification. Without them, the falsification time increases sharply by 8.1$\times$ (from 97.0s to 781.3s), and both the success rate and sub-specification violation coverage greatly decrease.
These results highlight the importance of the synthesis and refinement processes in achieving efficient and effective falsification.

\vspace{0.2em}
\noindent\textbf{w/o $\epsilon$-greedy Sampling.} \tool without the $\epsilon$-greedy sampling strategy (described in Section~\ref{subsec:sampling}). We directly use the conjunctive specification in falsification but not sample from it. \tool achieves an average sub-specification violation coverage of only 46.0\%, which is 37.9\% (from 83.9\% to 46.0\%) lower.
This result shows the importance of the $\epsilon$-greedy sampling in improving sub-specification coverage.

These results highlight the importance of each core process of \tool in achieving efficient and effective falsification.

\begin{table}[t!]
    \small
    \centering
    \caption{Average running time of each process in {\sf \textsc{Synthify}} on 8 control systems for 50 falsification trials.}
    \label{tab:overhead}
    \renewcommand{\familydefault}{\sfdefault}\normalfont
    \begin{tabular}{c|c|c}
        \hline\hline
        Synthesis & Refinement & Proxy Program \\
        \hline
        9.1s / 9.4\% & 1.4s / 1.4\%  & 4.2s / 4.3\%\\
        \hline \hline
        AI Controller & Plant Execution & Algorithm \\
        \hline
        5.0s / 5.2\% & 64.3s / 66.3\% & 13.0s / 13.4\%\\
        \hline\hline
    \end{tabular}
\end{table}

\subsection{Analysis of Running Time}

Table~\ref{tab:overhead} shows the average running time of each process in \tool
 on 8 control systems for 50 falsification trials.
The synthesis process takes only 9.1 seconds on average, accounting for 9.4\% of the total running time. The refinement process takes even less time, averaging 1.4 seconds, or 1.4\% of the total running time.
These results show that the synthesis and refinement are both lightweight and do not significantly impact the overall running time of \tool.
The execution of the proxy program and the AI controller take 4.3\% and 5.2\% of the total running time, respectively, neither of which is significant. This result clearly shows AI controller is no longer the bottleneck of the running time in falsification as described in Section~\ref{sec:motivation}, and the synthesis of a computationally cheaper proxy program has led to significant improvements in efficiency, shifting the bottleneck to the execution of the plant, which can be future work to further improve the efficiency of \tool.

\subsection{Alignment between Proxy Programs and AI Controllers}

\tool's effectiveness relies on the functional alignment between the synthesized proxy programs and AI controllers. Intuitively, the closer the proxy programs match the behavior of the AI controllers, the more likely they are to produce similar outputs, with fewer spurious violations during the refinement process—i.e., if the AI controllers produce unsafe outputs that result in safety violations, then the proxy programs are likely to do the same. This section conducts an analysis to evaluate this alignment. For this evaluation, we synthesize proxy programs for the 8 control systems using the same experiment settings as described in Section~\ref{sec:exp_setup}. Both the proxy programs and AI controllers run for 1000 time steps to gather their outputs. Additionally, we run the falsification for 50 trials and track the number of spurious violations encountered during refinement.

We first use the mean absolute error (MAE) to directly quantify the difference between the outputs of the AI controllers and the proxy programs. A lower MAE suggests stronger alignment. Since the AI controllers produce continuous outputs rather than discrete labels, the accuracy metric used in ~\cite{DBLP:conf/seke/WangHZSH21} for label-based predictions is not suitable for this comparison. MAE, which is commonly used to evaluate regression models~\cite{brassington2017mean}, is more appropriate in this context. To ensure comparability, we normalize the outputs of both the AI controllers and the proxy programs to the range [0, 1], as their scales differ across various control systems. After normalization, we calculate the MAE using the following formula:
\begin{equation}
    \text{MAE} = \frac{1}{N} \sum_{i=1}^{N} |y_{\text{AI}}^{(i)} - y_{\text{proxy}}^{(i)}|
\end{equation}
where $N$ represents the number of time steps, and $y_{\text{AI}}^{(i)}$ and $y_{\text{proxy}}^{(i)}$ denote the normalized outputs of the AI controller and proxy program at time step $i$, respectively. We repeat the experiments 10 times and present the average results in Table~\ref{tab:spurious}. The MAE for the outputs averages 0.19 across all control systems, indicating that the proxy programs' outputs differ by only 19\% (as the outputs are already normalized to the [0, 1] range) from those of the AI controllers. This difference is considered small and acceptable by the prior studies~\cite{haq2021can,haq2020comparing} that also evaluate the alignment between the outputs of real systems and their proxies. These results demonstrate a strong functional alignment between the proxy programs and the AI controllers.

Beyond comparing outputs, we also evaluate alignment between the AI controllers and proxy programs by examining the number of spurious violations during the refinement process, as shown in Table~\ref{tab:spurious}. Our results show that, on average, \tool identifies 2.1 spurious violations for the purpose of refining the proxy programs. The minimal occurrence of spurious violations further demonstrates the proxy programs' effectiveness in capturing the AI controllers' behavior. This strong alignment provides a solid foundation for applying falsification techniques on the proxy programs to identify safety violations in AI-enabled control systems.

\begin{table}[t!]
    \small
    \centering
    \caption{The mean absolute error (MAE) of the outputs of the AI controllers and the synthesized proxy programs, and the number of spurious violations used in the refinement. Results are averaged over experiments repeated 10 times.}
    \label{tab:spurious}
    \renewcommand{\familydefault}{\sfdefault}\normalfont
    \begin{tabular}{c|c|c}
    \hline\hline
    Control System & MAE (Outputs)  & \# Spurious Violations \\ \hline\hline
    CartPole       & 0.31   & 1.6  \\ \hline
    Pendulum       & 0.14  & 4.4                    \\ \hline
    Quadcopter     & 0.10   & 6.3            \\ \hline
    Self-Driving   & 0.07  & 3.5                  \\ \hline
    Lane Keeping   & 0.22  & 1.0                    \\ \hline
    4-Car Platoon  & 0.11   & 0                   \\ \hline
    8-Car Platoon  & 0.18   & 0                  \\ \hline
    Oscillator     & 0.40   & 0                    \\ \hline\hline
    Average        & 0.19  & 2.1                  \\ \hline \hline
    \end{tabular}
\end{table}

\subsection{Handling Large AI Controllers}

As mentioned in Section~\ref{sec:benchmarks}, the AI controllers used in our evaluations vary significantly in size, with neuron counts ranging from 440 to 1200, which are relatively large compared to those used in other studies like~\cite{zhang2022falsifai, tran2020nnv}, providing a diverse basis for evaluation. However, we anticipate that larger models will be integrated into control systems as hardware continues to advance. Therefore, we conduct additional experiments to evaluate \tool's performance on control systems with larger AI controllers. These new experiments use three variant models trained in the {\sf CartPole}, {\sf Pendulum}, and {\sf Self-Driving} control systems, also released by Zhu et al.~\cite{zhu2019inductive}, which feature larger AI controllers with 3 layers containing 1200, 900, and 800 neurons, respectively—an increase of 2.4$\times$ to 6.6$\times$ the number of neurons compared to the 8 controllers used in our main experiments.

All experiments are conducted under the same settings described in Section~\ref{sec:exp_setup}, and the results are presented in Table~\ref{tab:add1}. \tool consistently outperforms the baseline in both settings, achieving a relative improvement of 155.1\% in the falsification success rate across 50 trials and 14.5$\times$ more violations found within 10 minutes. Statistical tests confirm the significance and effect size of these improvements, with $p$-values below 0.01 and large effect sizes (except for the {\sf Self-Driving} system, where both our \tool and \baseline achieve a 100\% success rate). Additionally, sub-specification violation coverage is significantly improved compared to \baseline, with an average relative increase of 219.5\%, as shown in Table~\ref{tab:add2}. These results demonstrate that \tool remains highly effective at falsifying safety violations in control systems with large AI controllers, underscoring its potential for future applications in more complex systems.

\begin{table*}[t!]
    \footnotesize
    \centering
    \caption{Falsification results of {\sf \textsc{Synthify}} and {\sf \textsc{PSY-TaLiRo}} on three control systems with large AI controllers. Relative improvements are shown in parentheses. The $\hat{A}_{12}$ column shows the effect size, where $N$, $M$, and $L$ indicate negligible, medium, and large effect sizes, respectively.}
    \label{tab:add1}
    \renewcommand{\familydefault}{\sfdefault}\normalfont
    \scalebox{0.87}{
    \begin{tabular}{@{}c|c|c|c|c|c|c|c|c@{}}
    \hline
    \hline
    & \multicolumn{4}{c|}{Setting 1 (50 trials)}  & \multicolumn{4}{c}{Setting 2 (10 minutes)}  \\ \hline\hline
    System & \baseline & \tool           & {\it p}-value     & {\it \^A}\textsubscript{12} & \baseline & \tool                  & {\it p}-value     & {\it \^A}\textsubscript{12} \\ \hline
    CartPole (Large)   & 17.6\%            & \textbf{100.0\% (+468.2\%)}  & 7.49e-3\textless{}0.01  & {\it L} (1.00)     & 15.6          & \textbf{818.5 (52.5$\times$)}  & 1.83e-3\textless{}0.01 & {\it L} (1.00)     \\ \hline
    Pendulum (Large)   & 0.0\%             & \textbf{100.0\% (---)} & 3.91e-3\textless{}0.01 & {\it L} (1.00)     & 0.0               & \textbf{49.3 (---)}                 & 1.82e-5\textless{}0.01 & {\it L} (1.00)     \\ \hline
    Self-Driving (Large) & 100.0\%            & 100.0\%  & 1.0 & {\it N} (0.50)     & 65.0             & \textbf{298.9 (4.6$\times$)} & 9.34e-5\textless{}0.01 & {\it L} (1.00)     \\
    \hline \hline
    Average      & 39.2\%            & \textbf{100.0\% (+155.1\%)} & ---                    & ---            & 26.9             & \textbf{388.9 (14.5$\times$)}   & ---                    & ---            \\ \hline\hline
    \end{tabular}
    }
\end{table*}

\begin{table}[t!]
    \small
    \centering
    \caption{
        Sub-specification violation coverage of {\sf \textsc{Synthify}} and {\sf \textsc{PSY-TaLiRo}}. The $\hat{A}_{12}$ column shows the effect size where $N$ and $L$ indicate negligible and large effect size, respectively.}
    \label{tab:add2}
    \renewcommand{\familydefault}{\sfdefault}\normalfont
    \begin{tabular}{c|c|c|c|c}
        \hline\hline
    Control System   & \baseline & \tool & {\it p}-value     & {\it \^A}\textsubscript{12} \\ \hline \hline
    {\sf CartPole (Large)}     & 12.5\%            & \textbf{100.0\% (+700.0\%)}  & 3.9e-3\textless{}0.01 & {\it L} (1.00)     \\ \hline
    {\sf Pendulum (Large)}    & 0.0\%             & \textbf{100.0\% (---)}   & 3.7e-4\textless{}0.01 & {\it L} (1.00)     \\ \hline
    {\sf Self-Driving (Large)} & 50.0\%            & \textbf{100.0\% (+100.0\%)} & 3.9e-3\textless{}0.01 & {\it L} (0.80)     \\ \hline \hline
    Average      & 31.3\%            & \textbf{100.0\% (+219.5\%)}  & ---                    & ---            \\ \hline\hline
    \end{tabular}
\end{table}

\subsection{Hyperparameters of the Sampling Strategy}

Our $\epsilon$-greedy sampling strategy (Section~\ref{subsec:sampling}) is governed by the hyperparameter $\epsilon$, which controls the balance between selecting a sub-specification randomly (exploration) and selecting the most promising one (exploitation). To find the optimal value of $\epsilon$, we use grid search, a common method for hyperparameter tuning~\cite{bergstra2012random}. Grid search evaluates performance for each hyperparameter within a predefined set of options. In our experiments, we vary $\epsilon$ within the range [0.0, 1.0] in increments of 0.1, while keeping all other hyperparameters constant (i.e., following the settings described in Section~\ref{sec:exp_setup}). A value of 0.0 corresponds to a fully greedy strategy, always selecting the most promising sub-specifications, while 1.0 corresponds to a fully exploratory strategy, selecting sub-specifications randomly. We conduct 50 falsification trials and evaluate \tool's sub-specification violation coverage across 8 control systems. The results, shown in Table~\ref{tab:hyperparameters}, indicate that \tool's performance improves with increasing $\epsilon$ values, with an optimal value of 0.9 yielding the highest average sub-specification violation coverage of 93.3\%. Consequently, we set $\epsilon$ to 0.9 in our experiments by default. However, the optimal value may vary depending on the control system and safety specification, necessitating further tuning in practice.

\begin{table*}[t!]
    \footnotesize
    \centering
    \caption{Sub-specification violation coverage of {\sf \textsc{Synthify}} with different hyperparameters of the $\epsilon$-greedy sampling strategy.}
    \label{tab:hyperparameters}
    \renewcommand{\familydefault}{\sfdefault}\normalfont
    \scalebox{0.98}{
    \begin{tabular}{@{}c|c|c|c|c|c|c|c|c|c|c|c@{}}
    \hline
    \hline
    System & 0.0 & 0.1 & 0.2 & 0.3 & 0.4 & 0.5 & 0.6 & 0.7 & 0.8 & 0.9 & 1.0 \\ \hline \hline
    CartPole    & 50.0\%            & 75.0\% & 50.0\%  & 50.0\%    & 100\%            & 100.0\%    &100.0\%  & 100.0\%  & 100.0\% & 100.0\% & 100.0\%   \\ \hline
    Pendulum     & 25.0\%            & 25.0\%  &  25.0\%  &  25.0\%    &  25.0\%            &  25.0\%    & 25.0\%  &  25.0\%  &  25.0\% &  25.0\% &  25.0\%   \\    \hline
    Quadcopter  & 75.0\%            & 75.0\%  & 75.0\%   & 75.0\%    & 50.0\%             & 100.0\%    &100.0\%  & 100.0\%  & 100.0\% & 100.0\% & 100.0\%   \\ \hline
    Self-Driving & 25.0\%            & 25.0\% & 25.0\%  & 25.0\%    & 75.0\%            & 75.0\%    &100.0\%  & 75.0\%  & 100.0\% & 100.0\% & 100.0\%   \\ \hline
    Lane Keeping & 50.0\%            & 50.0\% & 50.0\%  & 50.0\%   & 50.0\%           & 100.0\%    &100.0\%  & 50.0\%  & 100.0\% & 100.0\% & 50.0\%   \\ \hline
    4-Car Platoon & 71.4\%    & 85.7\% & 92.9\%  & 92.9\%    & 92.9\%            & 100.0\%    &92.9\%  & 92.9\%  & 100.0\% & 100.0\% & 92.9\%   \\ \hline
    8-Car Platoon & 73.3\%  & 86.7\%  & 83.3\%  & 80.0\%  & 76.7\%  & 86.7\%  & 83.3\% & 86.7\% &  83.3\% & 90.0\%  & 76.7\%  \\ \hline
    Oscillator & 100.0\%            & 100.0\% & 100.0\%  & 100.0\%    & 100.0\%         & 100.0\%    &100.0\%  & 100.0\%  & 100.0\% & 100.0\% & 100.0\%   \\
    \hline \hline
    Average     & 59.8\%            & 68.3\% & 68.3\%  & 67.3\%    & 73.3\%            & 86.7\%    & 90.0\%  & 80.0\%  & 90.0\% & {\bf 93.3\%} & 83.3\%   \\ \hline\hline
    \end{tabular}
    }
\end{table*}

\subsection{Falsifying Intricate Safety Specifications}

\begin{table}[t!]
    \caption{Falsification results on {\sf Self Driving} control system with two cases of intricate safety specifications.}
    \label{tab:case_studies}
    \footnotesize
    \centering
    \renewcommand{\familydefault}{\sfdefault}\normalfont
    \scalebox{0.86}{
    \begin{tabular}{@{}c|c|c|c|c|c|c@{}}
    \hline
    \hline
    \multirow{2}{*}{Safety Specification} & \multicolumn{2}{c|}{Setting 1 (50 trials)} &  \multicolumn{2}{c|}{Setting 2 (10 minutes)} & \multicolumn{2}{c}{Coverage} \\ \cline{2-7}
    &   \baseline &  \tool & \baseline &  \tool & \baseline &  \tool \\ \hline\hline
    $\square_{[0,200]} \Diamond_{[0,20]}\left(|\eta|<90^{\circ} \wedge|d|<2.0\right)$  & 100.0\% & 100.0\% & 98.8  & {\bf 378.8 (3.8$\times$)}  & 50.0\% & {\bf 100.0\% (+100.0\%)} \\ \hline
     $\square_{[0,200]}\left(|d|>2.0 \rightarrow \Diamond_{[0,20]}(|d|<2.0)\right)$   & 66.7\% & {\bf 100.0\% (+49.9\%)} & 38.4 & {\bf 154.5 (4.0$\times$)} & 50.0\% & {\bf 100.0\% (+100.0\%)}  \\ \hline \hline
    Average & 83.4\% & {\bf 100.0\% (+19.9\%)} & 68.6 & {\bf 266.7 (3.9$\times$)} & 50.0\% & {\bf 100.0\% (+100.0\%)} \\ \hline\hline
    \end{tabular}
    }
\end{table}

In our experiments, the conjunctive safety specifications typically take the form of $\square_{\mathcal{I}} \varphi_1 \wedge \varphi_2$, requiring the system to satisfy both $\varphi_1$ and $\varphi_2$ at every time step within the interval $\mathcal{I}$. However, real-world safety specifications are often more intricate, involving nested temporal operators or complex logical expressions. We believe \tool is capable of falsifying such complex safety specifications, benefiting from our implementation on top of the infrastructure components from \baseline and utilizing the off-the-shelf tool RTAMT to compute the qualitative robustness semantics (as we mentioned in Section~\ref{sec:implementation}). Both can handle complex safety specifications, equipping \tool with the same capability.

To further quantitatively assess \tool's effectiveness in falsifying more intricate safety specifications, we construct additional specifications that involve nested temporal operators, such as $\square_{\mathcal{I}} \left(\Diamond_{\mathcal{I}} \varphi\right)$ or $\square_{\mathcal{I}} \left(\varphi_1 \rightarrow \Diamond_{\mathcal{I}} \varphi_2\right)$, for more experiments. Since the benchmarks used in our experiments only include specifications of the form $\square_{\mathcal{I}} \varphi$, these additional specifications can be only manually designed, requiring domain expertise and making them difficult to scale. We only evaluate \tool on two intricate safety specifications using the Self Driving control system introduced in Section~\ref{sec:motivation}, which is representative of real-world autonomous driving systems, as a proof of concept.

The safety specifications and corresponding results are presented in Table~\ref{tab:case_studies}. The first specification ($\square_{[0,200]} \Diamond_{[0,20]}\left(|\eta|<90^{\circ} \wedge|d|<2.0\right)$) requires the car to reach a state where the steering angle $\eta$ is within 90 degrees and the distance $d$ between the car's front and the road's centerline is within 2.0 meters, within 20 time steps at every point during the 200 time steps, enforcing a proactive condition where the system must regularly check and ensure safety. The second specification ($\square_{[0,200]}\left(|d|>2.0 \rightarrow \Diamond_{[0,20]}(|d|<2.0)\right)$) requires the car to steer back to a safe distance within 20 time steps after $d$ exceeds 2.0 meters during the 200 time steps, enforcing a reactive condition where the system must respond to unsafe situations. We evaluate \tool on these specifications using the same experimental settings as in Section~\ref{sec:exp_setup} and the same evaluation metrics. The results show that \tool is highly effective in falsifying these complex safety specifications, achieving a 100\% success rate across all case studies, which is 19.9\% higher than the baseline. Additionally, the number of uncovered violations increased by 3.9$\times$ within 10 minutes compared to the baseline. The sub-specification violation coverage also improved significantly, with an average relative improvement of 100\%. These results highlight \tool's ability to handle more complex safety specifications and its effectiveness beyond the simpler conjunctive specifications used in the primary experiments, demonstrating its potential for real-world applications.

\subsection{Potential Applications in End-to-End AI Systems}

The adoption of end-to-end AI-enabled control systems has recently surged in various domains, such as autonomous driving~\cite{10258330} and robotics~\cite{guo2023recent}. These systems use a single neural network to map raw sensor data directly to control signals, achieving enhanced performance, as exemplified by real-world systems like Tesla FSD~\cite{tesla} and NVIDIA DRIVE~\cite{NVIDIA}. We suggest that \tool has been developed in line with the trend of adopting end-to-end AI systems, and quantitatively evaluated through experiments on 8 end-to-end AI-enabled control systems. A further qualitative discussion supporting \tool's effectiveness in this context is provided below.

First and foremost, \tool enhances the efficiency and sub-specification coverage of falsification for end-to-end AI systems. As mentioned in Section~\ref{sec:intro}, these systems typically involve complex AI controllers with numerous parameters for processing inputs and generating control signals in one go~\cite{bojarski2016end,julian2016policy}. This complexity leads to high computational costs during testing~\cite{haq2022efficient,pang2022mdpfuzz}. \tool mitigates this by synthesizing efficient proxy programs that replace the AI controllers, significantly reducing the runtime of falsification. Moreover, thorough testing of end-to-end AI systems requires diverse test inputs~\cite{lou2022testing,10.1145/3597503.3639149}. \tool meets this need by maximizing sub-specification coverage of the conjunctive safety specification, as a more diverse set of test inputs is likely to falsify more sub-specifications~\cite{10.1145/3597926.3598108, viswanadha2021parallel}; intuitively vice versa. The $\epsilon$-greedy sampling strategy employed by \tool promotes the exploration of various sub-specifications, enabling the falsification process to uncover diverse safety violations by targeting different aspects of the safety requirements.

Furthermore, certain modules in \tool, such as our synthesis and $\epsilon$-greedy sampling strategy, can complement existing verification and testing tools for end-to-end AI systems. For instance, our synthesis module could be applied to generate simple programs that closely approximate the behavior of complex AI controllers but are more amenable to formal verification methods~\cite{fulton2020formal, 10.1145/3596444}, like constructing barrier certificates~\cite{zhu2019inductive} or performing reachability analysis~\cite{tran2019safety}. In the context of testing, our $\epsilon$-greedy sampling strategy can be integrated into existing falsification tools like~\cite{dreossi2019compositional,zhang2022falsifai} to strengthen their ability to falsify conjunctive specifications. Alternatively, it can support automated testing frameworks that handle multiple objectives~\cite{panichella2018testing,haq2022efficient}, helping developers efficiently prioritize and focus on the most promising objectives during testing.

Beyond testing and verification, \tool aids developers in inspecting end-to-end AI systems for interpretability and explainability—two essential factors in ensuring the trustworthiness of AI-enabled systems~\cite{glanois2024survey}. Interpretability focuses on understanding the system's inherent characteristics. Developers can leverage \tool's identified safety violations and corresponding falsifying inputs to derive formal specifications and properties of AI controllers~\cite{gopinath2019property,rozier2016specification}, serving as proof of the system's safety (or lack thereof). Explainability, on the other hand, involves using external proxies to clarify the system's operations~\cite{vouros2022explainable}. \tool's synthesized proxy programs, with straightforward control logic, can provide explanations for generated control signals by transparently revealing the execution process. This allows developers to better understand the AI controller's decision-making and identify potential vulnerabilities or biases within the system.

\subsection{Limitations and Potential Solutions}

While \tool demonstrates success in several cases within our evaluation, its primary limitation lies in generalizing to control systems with more complex inputs, such as images or videos. These inputs often have many dimensions and correspond to numerous variables in our linear program sketch, which can significantly expand the search space during the synthesis process, making it difficult to identify a suitable proxy program that accurately captures the AI controller's behavior. One potential solution is to use dimensionality reduction techniques like t-distributed Stochastic Neighbor Embedding (t-SNE)~\cite{NIPS2002_6150ccc6} to project the inputs into a lower-dimensional space before synthesizing the proxy program. Alternatively, we could move away from the current linear program sketch and instead synthesize a compact neural network capable of processing high-dimensional inputs using techniques like model compression~\cite{10.1145/3639475.3640097,10.1145/3551349.3556964,10.1145/3708525}. Despite this limitation, our methods are currently well-suited for control systems with numerical inputs like scalar values or vectors, thus we leave the extension to more complex inputs as future work.

\subsection{Threats to Validity}
\label{sec:threats}

One threat to internal validity is the randomness of the falsification algorithms used in our experiments. To mitigate it, we repeat each experiment 10 times and use appropriate statistical tests to account for both statistical significance and effect size, as recommended by Arcuri and Briand~\cite{arcuri2011practical}.

In terms of external validity, one threat is that the results of our analysis may not be generalizable. To mitigate it, we have carefully selected a representative set of control systems with diverse tasks to evaluate our method and baseline under different settings, which ensures that our results are not biased. However, as these control systems take the input as numerical values, the generalization of our results to control systems with more complex inputs, such as images or videos, may need further investigation.

One potential threat to construct validity is that the evaluation metrics may not fully capture the performance of falsification tools. To mitigate it, we used a total of five commonly-used evaluation metrics that are introduced in Section~\ref{sec:metrics} to compare the effectiveness and efficiency of \tool and the baseline from a comprehensive set of perspectives.

%% file: sections/rel_work.tex
\section{Related Work}
\label{sec:rel_work}

This section presents an overview of studies relevant to this paper, including (1) falsification of control systems, (2) program synthesis for AI models, and (3) surrogate-assisted testing and verification, and (4) multi-objective optimization in AI-enabled control systems.

\subsection{Falsification of AI-Enabled Control Systems}

Falsification~\cite{eddeland2020industrial,annpureddy2011s,adimoolam2017classification,menghi2020approximation,waga2020falsification,ernst2021falsification,tuncali2018experience,chandratre2023stealthy} has emerged as an alternative to formal verification that provides better efficiency to uncover violations of a given safety specification. There comes many falsification techniques supported by optimization algorithms, such as Hill Climbing~\cite{mathesen2021efficient}, Simulated Annealing~\cite{abbas2012convergence,thibeault2021psy}, Monte Carlo Tree Search~\cite{zhang2018two,zhang2021effective}, gradient-based methods~\cite{abbas2014functional,yaghoubi2017hybrid}, as well as some machine learning-guided methods~\cite{aineto2023falsification,zhang2021figcps}.

To date, there has been limited attention directed toward the falsification of AI-enabled control systems, with only a few early attempts~\cite{yaghoubi2019gray,song2022cyber,zhang2022falsifai,dreossi2019verifai,xie2023mosaic}. Among these studies, Song et al.~\cite{song2022cyber} evaluate the reliability of AI-enabled control systems using two existing falsification tools~\cite{annpureddy2011s,donze2010breach} and highlight that they need further improvements, which necessitates our work. Zhang et al.~\cite{zhang2022falsifai} and Yaghoubi et al.~\cite{yaghoubi2019gray} develop gray-box falsification techniques that make different assumptions compared to our study. They assume the AI controllers or plants accessible and collect not only the input and output signals from the control system but also intermediate information of the AI controllers or plants as guidance for falsification. In comparison, our approach is a black-box falsification tool that solely relies on the input and output signals, making it more practical particularly in real-world applications where the controller or plants can be complex or even inaccessible if they are confidential products from third-party vendors.

VerifAI~\cite{dreossi2019verifai}, NNV~\cite{tran2020nnv}, and Mosaic~\cite{xie2023mosaic} are all safety analysis frameworks for AI-enabled control systems, each incorporating a falsification module. However, there are key distinctions between these tools and our \tool. VerifAI is designed for systems with AI models in the sensor components, while our work primarily concerns systems with AI models in the controller parts. In addition, both NNV and Mosaic require specific input specifications that differ from those supported by our \tool. NNV relies on assert statements as safety specifications, while Mosaic necessitates probabilistic computation tree logic (PCTL) specifications. These particular specifications are not compatible with our benchmark systems, and our \tool, like many other existing falsification tools, operates based on Signal Temporal Logic (STL) specifications~\cite{thibeault2021psy,annpureddy2011s,donze2010breach}. Attempting to translate STL specifications into assert statements or PCTL specifications would introduce additional manual efforts, which are not only time-consuming but also prone to errors. Such translations also fall outside the scope of this paper. Consequently, in our comparison experiments, we have chosen to use \baseline~\cite{thibeault2021psy} as our baseline tool.

\subsection{Surrogate-Assisted Falsification}

Several studies have also focused on the safety analysis of control systems, though not specifically AI-enabled ones, and aligned with our work by leveraging surrogate models. Some of these studies approximate control systems with surrogate models trained on data collected from the systems, which can replace the original systems during falsification. For example, Menghi et al.~\cite{menghi2020approximation} propose ARIsTEO, which replaces computationally-intensive control systems with surrogate models derived from system identification techniques to improve falsification efficiency. Waga et al.~\cite{waga2020falsification} introduce FalCAuN, which learns a Mealy machine as a surrogate model and falsifies it using model checking, while Bak et al.~\cite{bak2024falsification} learn a Koopman model and falsify it through reachability analysis. Atanu et al.~\cite{10.1145/3641399.3641401} propose to build a DNN model that approximates the behavior of the control system and use adversarial attack algorithms for falsification.

Another set of studies integrates surrogate models into the falsification process. For instance, several studies~\cite{10.1145/3126521,ramezani2022falsification,moss2023bayesian,zhang2021gaussian,qin2022statistical} employ Bayesian optimization with Gaussian processes as surrogate models to estimate the robustness semantics of STL specifications, guiding the search for safety violations. Additionally, Peltom{\"a}ki et al.~\cite{peltomaki2022falsification} propose training a surrogate DNN model to directly generate falsifying inputs for control systems.

While our synthesized proxy programs may resemble surrogate models, they fundamentally differ from those in above studies. First, in terms of form, our proxy programs implement linear controllers, whereas surrogate models in prior work are typically DNNs (e.g.,~\cite{10.1145/3641399.3641401, peltomaki2022falsification}) or machine learning models (e.g.,~\cite{10.1145/3126521,qin2022statistical}). The simplicity of our programs offers greater efficiency. Second, regarding purpose, our proxy programs replicate the functionality of AI controllers, acting as their stand-ins to facilitate efficient falsification by minimizing the need to run the controllers themselves. This addresses the unique computational challenges associated with AI controllers, which are expensive to execute due to their vast number of parameters. In contrast, surrogate models in earlier work are usually designed to replace control systems or predict robustness semantics. Third, in terms of underlying technique, our approach utilizes Evolution Strategy for program synthesis, while prior methods often rely on Bayesian optimization (e.g.,~\cite{qin2022statistical, ramezani2022falsification}) or DNNs (e.g.,~\cite{menghi2020approximation, 10.1145/3641399.3641401,peltomaki2022falsification}) to build surrogate models.

These distinctions not only make our approach unique but also allow it to complement existing methods, further enhancing the efficiency of falsification in AI-enabled control systems. While combining our approach with existing surrogate-assisted falsification methods is a promising direction, a detailed evaluation is beyond the scope of this work and is left for future research.

\subsection{Program Synthesis for AI models}

Although program synthesis has been studied for decades~\cite{sobania2022comprehensive}, very little attention has been paid to program synthesis specialized for AI models. A few studies~\cite{verma2018programmatically,trivedi2021learning} have proposed methods for generating programmatic policies from well-trained DRL models, which are more interpretable than neural networks. Additionally, Zhu et al.~\cite{zhu2019inductive} introduce an inductive synthesis-based method that can synthesize deterministic and verified programs from DRL-based models. Inala et al.~\cite{inala2020synthesizing} and Qiu et al.~\cite{qiu2022programmatic} synthesize programmatic controllers to perform continuous control tasks. Our approach differs from these studies in that we propose an Evolution Strategy-based algorithm that can effectively synthesize programs as proxies for AI controllers, which we believe is both technically novel and offers a unique perspective on improving the efficiency of falsification in AI-enabled control systems.

\subsection{Multi-Objective Optimization in AI-Enabled Control Systems}

Our focus on covering various sub-specifications can be viewed as a multi-objective optimization problem. Thus, a relevant body of work lies in the multi-objective optimization within AI-enabled control systems, and we particularly mention those in the areas of testing, repair, and enhancement.

In the context of testing, Zolfagharian et al.~\cite{zolfagharian2023search} and Haq et al.~\cite{haq2022efficient,haq2023many} propose multi-objective search-based testing methods for AI-enabled control systems, accounting for diverse fitness functions or safety requirements. Further studies address repairing these systems. Lyu et al.~\cite{lyu2023autorepair} repair AI controllers' unsafe behaviors by adding patches to control signals via multi-objective optimization, while MORTAR~\cite{wang2024mortar} replaces unsafe signals from AI controllers with repaired alternatives through gradient-based optimization. Both methods solve the multi-objective problem of maximizing satisfaction of safety requirements while minimizing changes to the original control signals. Regarding safety enhancement without directly repairing the violations, SIEGE~\cite{10144351} stands out by presenting an ensemble framework that synergistically combines multiple AI controllers to improve the system's reliability across multiple control objectives. GenSafe~\cite{zhou2024gensafe} corrects the behaviors of AI controllers that potentially deviate from safety constraints during training to enhance their safety while also maintaining performance at a satisfactory level. In addition to all of the above, Zhou et al.~\cite{10.1145/3639477.3639740} establish a benchmark for multiple robotic manipulation tasks, providing a platform to evaluate the safety and effectiveness of AI controllers in a multi-objective context.

%% file: sections/conclusion.tex
\section{Conclusion and Future Work}
\label{sec:conclusion}
In this study, we propose \tool, a novel falsification framework specifically designed for AI-enabled control systems. The key features of \tool include the use of a synthesized program to serve as a proxy of the AI controller, and the application of the $\epsilon$-greedy strategy to sample a promising safety specification from the conjunctive requirement. We evaluated \tool on 8 publicly-available control systems and compared it with a state-of-the-art falsification tool \baseline.
Our evaluation results show that \tool outperforms the baseline in terms of both effectiveness and efficiency.
We evaluated \tool on 8 publicly-available control systems and compared it with a state-of-the-art, open-source falsification tool \baseline.
These promising results demonstrate the high effectiveness and efficiency of \tool in the falsification of AI-enabled control systems.

In the future, we plan to extend \tool to AI-enabled control systems that take more complex inputs, such as images~\cite{9825775} and audio~\cite{9609154}. We also plan to integrate \tool with other safety analysis techniques like runtime shields~\cite{shi2024synthesizing} and bug localization~\cite{9678546,9825855} to further improve the efficiency of falsification in AI-enabled control systems. Furthermore, with the advent of Large Language Models (LLMs), investigating the use of code generation LLMs~\cite{9825794,yang2024acecoder,10.1145/3597503.3639074,10735776} for synthesizing proxy programs is a promising avenue for further research.

\ans{\noindent\textbf{Replication Package.} For reproducibility, we have made our code and data publicly available: \url{https://github.com/soarsmu/Synthify}.}